\documentclass[12pt]{article}
\usepackage{jheppub}
\usepackage{amsfonts,ulem,bbold}   
\usepackage{amssymb,amsmath,hyperref}
\usepackage{xcolor}
\usepackage{feynmp}
\usepackage{multirow}
\usepackage{scrextend} %needed for double reference to a footnote
\usepackage{tablefootnote} % needed for footnote in table

\def\lsim{\mathrel{\rlap{\lower3pt\hbox{\hskip0pt$\sim$}}
   \raise1pt\hbox{$<$}}}         %less than or approx. symbol
\def\gsim{\mathrel{\rlap{\lower4pt\hbox{\hskip1pt$\sim$}}
   \raise1pt\hbox{$>$}}}         %greater than or approx. symbol

\def\l{\left}
\def\r{\right}
\def\fr{\frac}
\def\la{\label}
\def\d{\partial}
\newcommand{\p}{\bar{P}}  
 
\def\be{\begin{equation}}
\def\ee{\end{equation}}
\def\ba{\begin{eqnarray}}
\def\ea{\end{eqnarray}}	
\def\vphi{\varphi}

\newcommand{\sfrac}[2]{{\textstyle\frac{#1}{#2}}}
\newcommand{\vp}{ \vec{\eta}}

\title{Zoology of condensed matter:\\
Framids, ordinary stuff, extra-ordinary stuff} 

\author[a]{Alberto Nicolis,}
\author[a]{Riccardo Penco,}
\author[abc]{Federico Piazza,}
\author[d]{and Riccardo Rattazzi,}

\affiliation[a]{Physics Department and Institute for Strings, Cosmology, and Astroparticle Physics,\\
  Columbia University, New York, NY 10027, USA}

\affiliation[b]{Paris Center for Cosmological Physics and Laboratoire APC, \\
Universit\'{e} Paris 7, 75205 Paris, France}

\affiliation[c]{CPT, Aix Marseille Universit\'e, UMR 7332, 13288 Marseille,  France.}

\affiliation[d]{Institut de Th\'eorie des Ph\'enom\`enes Physiques, \\
EPFL Lausanne, Switzerland}

\abstract{
We classify condensed matter systems in terms of the spacetime symmetries they spontaneously break. In particular, we characterize condensed matter itself as any state in a Poincar\'e-invariant theory that spontaneously breaks Lorentz boosts while preserving at large distances some form of spatial translations, time-translations, and possibly spatial rotations. 
Surprisingly, the simplest, most minimal system achieving this symmetry breaking pattern---the {\it  framid}---does not
seem to be realized in Nature. Instead, Nature usually adopts a more cumbersome strategy: that of introducing {\it internal} translational symmetries---and possibly rotational ones---and of spontaneously breaking them along with their space-time counterparts, while preserving unbroken diagonal subgroups. This symmetry breaking pattern describes the infrared dynamics of ordinary solids, fluids, superfluids, and---if they exist---supersolids.
A third, ``extra-ordinary", possibility involves replacing these internal symmetries with other symmetries that do not commute with the Poincar\'e group,
 for instance the galileon symmetry, supersymmetry or gauge symmetries. Among these options,  we pick the systems based on the galileon symmetry, the ``{\it galileids}",  for a more detailed study. 
Despite some similarity, all different patterns produce truly distinct physical systems with different observable properties. For instance, the low-energy $2\to 2$ scattering amplitudes for the Goldstone excitations in the cases of framids, solids and galileids scale respectively as $E^2$, $E^4$, and $E^6$. 
Similarly the energy momentum tensor in the ground state is ``trivial" for framids ($\rho +p=0$), normal for solids ($\rho+p>0$) and even inhomogenous for galileids.
}

\begin{document}

\begin{fmffile}{graphs}

\maketitle
\flushbottom

%%%%%%%%%%%%%%%%%%%%%%%%%%%%%%%%%%%%%%%%%%%%
%%%%%%%%%%%%%%%%%%%%%%%%%%%%%%%%%%%%%%%%%%%%
\section{Introduction}

When we think about condensed matter, we rarely invoke relativity as a guiding principle. The reason is twofold.
On the one hand, ordinary condensed matter systems in the lab are extremely non-relativistic: their mass density is  much bigger than their energy density and pressure, the propagation speeds of their excitations (e.g., phonons) are extremely subluminal, etc. On the other hand, and more to the point, each such system has an associated rest frame, which breaks the equivalence of all inertial frames and makes relativistic considerations apparently useless.  As a result, the collective excitations of macroscopic bodies are usually modelled with Lagrangians and Hamiltonians that have nothing to do with relativity.

However, sometimes it can be useful to keep in mind that---to the best of our knowledge---the fundamental laws of physics are Lorentz invariant, and that real-world condensed matter systems emerge as particular Lorentz-violating states subject to such  fundamentally relativistic laws. In other words, condensed matter systems break Lorentz invariance {\it spontaneously}. As Goldstone's theorem and its subsequent refinements (current algebra techniques, effective field theory) have taught us, spontaneous symmetry breaking  can have profound physical implications. For instance, in the case at hand, the statement that a superfluid's phonons have to non-linearly realize the spontaneously broken Lorentz boosts, forces their interactions to have a very constrained structure, involving solely powers of the combination
\be
\dot \pi +\sfrac12 \dot \pi^2 -\sfrac12 (\vec \nabla \pi)^2
\ee 
in the low-energy limit \cite{Son:2002zn}\footnote{The point is often made that for most condensed matter systems---which are non-relativistic in the first sense spelled out above---the relevant spontaneously broken boosts one should focus on are Galilei's rather than Lorentz's, in which case the invariant combination becomes $\dot \pi - \sfrac12 (\vec \nabla \pi)^2$   \cite{WittenWilczek}. However, Galilean relativity is certainly an excellent {\it approximation} to Lorentzian relativity in many physical situations, but it is never  more {\it precise} than the latter. So, if technically feasible, we see no harm in imposing full Lorentz invariance and just neglecting $(v/c)^2$ relativistic corrections when desired and appropriate.}.
This is much less general than what one would have for a generic superfluid in a fictitious world with no fundamental Lorentz invariance, where all  combinations of $\dot \pi$ and $(\vec \nabla \pi)^2$ would be allowed.
Likewise, for solids, spontaneously broken Lorentz invariance forces the phonons to appear in the action at low energies always through the particular combination \cite{Dubovsky:2005xd, Nicolis:2013lma}
\be
\nabla^i \pi^j + \nabla ^j \pi^i - \dot \pi^i \dot \pi^j + \vec \nabla \pi^i \cdot \vec \nabla \pi^j  \; .
\ee

In this paper, we take spontaneously broken Lorentz invariance as the {\it defining feature} of condensed matter.
The other symmetries that we postulate are unbroken spatial homogeneity---which in certain systems like solids is recovered only upon coarse-graining on large enough scales---and time-translational invariance. To make our (and the reader's) life easier, we focus on systems that also feature unbroken three-dimensional rotations, at least on large enough scales. We thus give up describing the peculiarities of anisotropic systems like crystalline solids, although extending our considerations and results in that direction is, if algebraically tedious, conceptually straightforward.

We leave open the possibility that the unbroken translational and rotational symmetries featured by a given condensed matter system---those governing the collective excitations, or quasi-particles---may not be those originally appearing in the Poincar\'e group. Rather, they can be linear combinations of the latter and of certain additional symmetries. 
As we will see, this seemingly exotic possibility is in fact ubiquitous in Nature, so much so that  we are not aware of any condensed matter system that does {\it not} implement it: all condensed matter systems seem to require some additional symmetries. 

As we will explain below, an hypothetical system without such symmetries would not have the usual degrees of freedom associated with the positions of infinitesimal volume elements, like ordinary solids and fluids, but only the degrees of freedom associated with the local rest frame picked by the system. In other words, for ordinary fluids and solids we can think of each volume element as having some position {\it and} some velocity, while for this system the position degree of freedom is absent---the `volume element' language itself is absent---and one can only talk about the local velocity of the medium. 
% We will be more explicit about this distinction in the main text. 
We dub such an hypothetical system {\it (type I) framid}, since it involves  the most economical set of ingredients that an homogeneous physical system needs to `pick a frame'. Its only degrees of freedom are the components of the velocity vector of the local rest frame of the medium. 

To be concrete, consider for example a relativistic theory featuring a homogeneous and isotropic state $| \psi \rangle$ that  breaks Lorentz boosts via a non-trivial expectation value for some four-vector local operator in the theory:
\be
\langle A_\mu (x) \rangle = \delta_\mu^0 \; .
\ee
Let us assume further that the original spacetime translations and spatial rotations appearing in the Poincar\'e group are unbroken by $| \psi \rangle$, meaning that there are no expectation values of local operators  breaking them. Then, according to our characterization above, this state describes a framid. The only local degrees of freedom whose presence is guaranteed by symmetry are the Goldstone excitations, which can be thought of as localized infinitesimal boosts of the order parameter:
\be
A_\mu(x) = \big(e^{i \vec \eta(x) \cdot \vec K}  \big)_\mu {}^\alpha \, \langle   A_\alpha (x)  \rangle \; ,
\ee
where $\vec \eta(x)$ denotes a triplet of Goldstone fields---the `framons'---and $\vec K$ is the vector of boost generators.
Since the medium does not break translations or rotations, it cannot be translated, rotated, stretched, compressed, twisted, or ``deformed'' in any standard spatial sense. It can only be boosted.

Now contrast this with the field-theoretical description of a solid for instance. To keep track of the positions of the individual volume elements, one needs to introduce a triplet of scalar fields $\phi^I(\vec x, t)$ ($I=1,2,3$), which can be thought of as giving the comoving (Lagrangian) coordinates of the volume element occupying physical position $\vec x$ at time $t$. The ground state of the system (at some reference external pressure) has
\be
\langle \phi^I(x) \rangle = x^I \; .
\ee
That is, each volume element is at rest and occupies its own equilibrium position. Such expectation values break Lorentz boosts of course, as desired, but they also break spatial translations and rotations.
To recover the observed homegeneity and isotropy of a solid at large scales, one {\it needs} to impose some internal translational and rotational symmetries acting on the fields,
\be
\phi^I \to \phi^I + a^I \; , \qquad \phi^I \to SO(3) \cdot \phi^I \; ,
\ee
so that the expectation values above preserve suitable linear combinations of spatial symmetries and internal ones.
As a result of breaking spatial translations and rotations, the solid, unlike the framid, admits standard spatial deformation degrees of freedom, parameterized by the phonons, which serve as Goldstone bosons for all the broken symmetries (including boosts.)

One might wonder whether the extra structure needed to describe ordinary solids and fluids in effective field theory---the additional internal symmetries---just corresponds to a suboptimal, redundant description of these systems. Is it possible that the framid simply corresponds to a more economical description of the same systems, rather than to a physically different system altogether? Is there perhaps a complicated field redefinition that maps the effective field theory of a framid into that of a solid or a fluid? In fact, the standard hydrodynamical description of a fluid never involves explicitly the individual volume elements' positions, but rather the energy density $\rho$, the pressure $p$, the fluid's four-velocity $u^\mu$, etc.---none of which breaks translations or rotations for a fluid at equilibrium, but only boosts. To settle the question, one should compute a physical observable and compare the answers one gets in the two cases. In sect.~\ref{sec_amplitude} we show that the $2 \to 2$ scattering amplitude for the Goldstone excitations at low-energies scales like $E^4$ in solids and fluids, and like $E^2$ in a framid, thus proving that the framid is a physically distinct  system rather than just a rewriting of solids and fluids. Since the framid corresponds to the most economical way to break Lorentz boosts spontaneously while preserving  homogeneity, isotropy, and time-translational invariance, it is surprising that Nature never uses it. We elaborate on possible reasons for this in sect.~\ref{why not}. We have no definite answer yet, but we identify  one stark feature that sets framids apart from ordinary stuff: the energy momentum tensor on their ground state is proportional to a cosmological constant term. From a condensed matter perspective such energy momentum is equivalent, by a tuning of the cosmologial term, to $\rho=p=0$. This property remarkably corresponds to the absence, in opposition to ordinary stuff, of a moduli space of homogeneous and isotropic vacuum solutions that can be associated with a change of boundary conditions, e.g. a change of pressure.

%Perhaps adding to the confusion is the fact that Nature seems to have no problem with {\it admixtures} of ordinary media and framids: we argue that the B-phase of superfluid helium 3 is described in our language by a superfluid interacting with a framid.
%

Beyond the simple framid and beyond ordinary condensed matter, there finally  are ``extra-ordinary" systems. These  break spacetime translations and possibly spatial rotations, but make up for them via extra symmetries that do {\it not} commute with the Poincar\'e group. Extra-ordinary systems form a possibly wide class whose thorough exploration we leave for future work.
In sect.~\ref{sec:galileids} we limit our discussion to a few representatives including the simplest ones,  the {\it galileids}. The latter are based on
%that we shall not investigate in full detail. We shall however focus on what seems like 
%At the opposite end of the spectrum of possible symmetry breaking patterns lie the {\it galileids}. These are systems that---like ordinary condensed matter---break spacetime translations and possibly spatial rotations, but that---unlike ordinary condensed matter---make up for them via extra symmetries that do {\it not} commute with the Poincar\'e group. The simplest example is that of
 a galileon field~\cite{Nicolis:2008in}, that is a scalar field $\phi(x)$ whose dynamics enjoy a generalized shift symmetry
\be \la{galshift}
\phi(x) \to \phi(x) + c + b_\mu x^\mu \; ,
\ee
where $c$ and $b_\mu$ are constant transformation parameters. 
At lowest order in derivatives, its equation of motion is a non-linear algebraic equation for its second derivatives, which admits a continuum of solutions of the form
\be
\phi(x) = A \, |\vec x|^2 + B \, t^2 \; ,
\ee
where $A$ and $B$ are suitable constants. Such a solution breaks Lorentz boosts as well as spacetime translations, but the latter can be made up for by the generalized shift symmetry (\ref{galshift}). That is, there is an unbroken linear combination of spacetime translations and shifts of $\phi$ which can serve as the symmetry defining homogeneity and time-translational invariance in the Lorentz-violating background above.
%In sect.~\ref{sec:galileids} we discuss possible generalizations of this mechanism, as well as the plausibility that it is actually at work in some condensed matter system.

For all these systems, we will only deal with the infrared degrees of freedom that are forced to be there by the symmetries---the Goldstone excitations. In particular, we will not discuss fermionic excitations, which are of course responsible for much of the interesting phenomenology of condensed matter systems in the lab.
With this qualification in mind, we want to classify all possible low-energy, long-distance dynamics of condensed matter. Then, our problem naturally splits into two questions:
\begin{enumerate}
\item
What are all the possible ways of breaking the Poincar\'e group and additional symmetries down to spatial translations, time-translations, and rotations (and possibly residual internal symmetries)? As mentioned above, it should be kept in mind that the unbroken translations and rotations can differ in general from those appearing in the Poincar\'e group. In other words, the breaking can `mix' some of the Poincar\'e symmetries with the additional ones.
\item
For each symmetry breaking pattern, what is the most general effective field theory governing the low-energy, long-distance dynamics of the associated Goldstone bosons?
\end{enumerate}
The first question is purely mathematical in nature, and is answered in the next section. The rest of the paper is devoted to answering the second.

%%%%%%%%%%%%%%%%%%%%%%%%%%%%%%%%%%%%%%%%%%%%
%%%%%%%%%%%%%%%%%%%%%%%%%%%%%%%%%%%%%%%%%%%%
\section{Classification of symmetry breaking patterns}\label{SSB}

We are interested in classifying all the  symmetry breaking patterns that can be associated with a static, homogeneous, and isotropic medium in a relativistic theory. We will thus assume that the full symmetry group of our system is made up of the Poincar\'{e} group, whose generators are 
\begin{align}
P_0 &\qquad \mbox{(time traslations)} \\
P_i & \qquad\mbox{(spatial traslations)} \\
J_i & \qquad\mbox{(rotations)} \\
K_i & \qquad\mbox{(boosts)}
\end{align}
and (possibly) of some additional {\it internal} symmetries---i.e. symmetries whose generators commute with the Poincar\'{e} generators listed above. 
[We will moreover assume the existence of a set of translation and rotation generators that govern the excitations inside the condensed matter system, and, in particular, that leave the ground state invariant,
\be
\bar{P}_0, \quad \bar{P}_i, \quad  \bar{J}_i  \qquad \mbox{(unbroken)}\, .
\ee
The above generators need not be the original ones appearing in the Poincar\'e group but they must obey the same algebra,
whose only non-vanishing commutators~are
\be \label{unbroken algebra}
[ \bar{J}_i , \bar{J}_j ] = i \epsilon_{ijk} \, \bar{J}^k , \qquad \qquad [ \bar{J}_i , \bar{P}_j ] = i \epsilon_{ijk} \, \bar{P}^k \; ,
\ee
or else there is no sense in which we can say that they generate translations and rotations. In the usual condensed matter jargon, $\bar{P}_0$ is the (usually, non-relativistic) Hamiltonian of the {\it quasi-particles} or {\it collective excitations} of the system.]

Clearly, this structure can be complicated at will by the addition of internal symmetries, both broken and unbroken. %However, we will require that all {\it unbroken} additional symmetries commute with the unbroken translations and rotations because, once again, otherwise we could not interpret $\bar{P}_0, \bar{P}_i$ and $\bar{J}_i$ as generators of ``effective'' space-time transformations {\bf{\color{red}---why not??---}}. This means that the inclusion of more unbroken symmetries is just a trivial generalization, unless there are also {\it broken} symmetries that fail to commute both with the additional unbroken symmetries and with unbroken translations and rotations. 
In general, there will be additional Goldstone modes associated with the broken symmetries, and they will transform linearly under all unbroken symmetries. However, one should keep in mind that there can be subtleties in the Goldstone phenomenon whenever broken symmetries do not commute with the unbroken $\bar P$'s~\cite{Low:2001bw}. For instance, some of the Goldstone excitations can acquire a gap \cite{Nicolis:2011pv, Nicolis:2012vf,Kapustin:2012cr,Watanabe:2013uya} and thus become irrelevant at low enough energies, whereas others may be removed altogether from the spectrum~\cite{Nicolis:2013sga} by imposing certain conditions known as {\it inverse-Higgs constraints}~\cite{Ivanov:1975zq}. Often these constraints can be interpreted as gauge fixing conditions that eliminate a redundancy in the parametrization of the Goldstone excitations; for certain systems though, this interpretation is not available and imposing inverse Higgs constraints simply amounts to integrating out gapped modes \cite{McArthur:2010zm,Nicolis:2013sga,Endlich:2013vfa}.  Regardless of their interpretation, the criterion for {\it when} inverse Higgs constraints can be imposed goes as follows: whenever the commutator between some unbroken translation $\bar P$ and a multiplet of broken generators $Q$ contains another multiplet of broken generators $Q'$, i.e. 
\be
[\bar{P}, Q ] \supset Q',
\ee
one can impose some inverse Higgs constraints and solve them to express the Goldstones of $Q$ in terms of derivatives of those of $Q'$. By doing so, one obtains another nonlinear realization of the same symmetry breaking pattern with fewer Goldstone fields.

\subsection{The eightfold way}

In light of these remarks, we propose to classify condensed matter systems based on which (if any) of the $\bar P$'s and $\bar J$'s involve internal symmetries. We find that there are in principle eight possible scenarios. For six of them there is the option to realize them purely with internal symmetries, while the other two necessarily require additional symmetries that do not commute with Poincar\'e. For each of these scenarios we are going to discuss the most minimal implementations---i.e. those that feature the {\it minimum} number of Goldstone excitations. If we denote all additional symmetry generators by `$Q$' (possibly with indices and other typographical appendages), the eight conceivable scenarios for static, homogeneous and isotropic condensed matter systems are: 
\begin{enumerate}
\item $\bar{P}_0 = P_0, \quad  \bar{P}_i =P_i, \quad \bar{J}_i = J_i$.\\
This first case is the most minimal scenario, in that it does not require any additional symmetry beyond the Poincar\'{e} group. The only space-time symmetries that are broken are the three Lorentz boosts, and thus we expect three Goldstone bosons.   We will call {\it type-I framid} a medium described by this symmetry breaking pattern. As already discussed in the Introduction, the simplest order parameter that realizes this scenario is a single vector operator that acquires a vev $\langle A_\mu (x) \rangle = \delta_\mu^0$.\footnote{Upon coupling to gravity, a type I framid gives rise to a Lorentz-violating modification of general relativity known as Einstein-\ae ther theory~\cite{Jacobson:2000xp}.}

\item $\bar{P}_0 = P_0 + Q, \quad  \bar{P}_i =P_i, \quad \bar{J}_i = J_i$. \\
%The commutation relations (\ref{unbroken algebra}) require that 
%%
%\be
%[ Q, P_i] = [ Q, J_i] = 0.
%\ee
%%
In the absence of additional symmetries, $Q$ is simply the generator of an internal $U(1)$ symmetry. This is the pattern of symmetry breaking associated with ordinary superfluids, which we will also call {\it type-I superfluids}. In this case, we know~\cite{Son:2002zn} that we can make do with only one Goldstone boson---the superfluid phonon---even though there is a total of four broken generators ($Q$ and the $K_i$'s). This is because $[\bar P_i, K_j] = i \delta_{ij} (Q - \bar P_0)$, and thus one can impose three inverse Higgs constraints and express the boost Goldstones in terms of the Goldstone of $Q$~\cite{Nicolis:2013lma}. 

Physically, the possibility of having  a single Goldstone mode follows from the fact that one can realize the SSB pattern above with a single weakly coupled scalar---the superfluid ``phase'' field---with a time-dependent vev, $\langle \psi (x) \rangle = t $. In this case,  $Q$ is realized as a shift-symmetry on the phase, $\psi \to \psi+a$. Equivalently, one can think of a weakly coupled complex scalar $\Phi(x)$ acted upon by $Q$ in the usual way, $\Phi \to e^ {ia} \Phi$, acquiring a `rotating' vev $\langle \Phi (x) \rangle = e^{i t}$. If $\Phi$ has a $U(1)$-invariant potential, the radial mode is gapped while the  angular mode is gapless and can be identified with the superfluid phonon.

The spectrum of Goldstone bosons for this scenario has been extensively studied also in the presence of an arbitrary compact group of internal symmetries~\cite{Watanabe:2011ec,Nicolis:2013sga}. Notice that any compact group larger than $U(1)$ inevitably leads---if broken---to additional Goldstone modes. This is because only the modes corresponding to broken generators that do not commute with $Q$ can in principle be eliminated by the inverse Higgs mechanism. However, it can be shown that {\it (i)} for a compact group one can always choose a basis of generators such that all non-commuting generators come in pairs~\cite{Nicolis:2012vf}, and that {\it (ii)} one can eliminate at most one Goldstone for each pair while keeping all non-linearly realized symmetries intact~\cite{Nicolis:2013sga}.  

Interestingly, additional Goldstone modes are {\it not} compulsory if one embeds $U(1)$ in a non-compact group. In fact, one can even add an infinite number of broken internal symmetries by promoting the $U(1)$ shift symmetry to internal (monotonic) diffeormorphisms,
\be \label{khronon}
\psi \to f(\psi) \; ,
\ee
 and still have a single Goldstone. A field enjoying such an internal symmetry arises for instance in the infrared limit of Ho\v{r}ava-Lifshitz gravity~\cite{Horava:2009uw,Blas:2009yd} and has been dubbed {\it khronon} in the gravity/cosmology literature~\cite{Creminelli:2012xb}.

\item $\bar{P}_0 = P_0, \quad  \bar{P}_i =P_i+ Q_i, \quad \bar{J}_i = J_i$.\\
The commutation relations (\ref{unbroken algebra})  require  
\be \la{3}
[ J_i, Q_j] = i \epsilon_{ijk} Q^k \; ,
\ee
which implies the $Q_i$'s cannot generate an internal symmetry. 
%$Q_i$ must be part of a non-trivial Lorentz multiplet, like a 4-vector $Q_\mu\to Q_i $, or an antisymmetric 2-index tensor $Q_{\mu\nu} \to \epsilon_{ijk}Q_{jk}$.
However, as we will see in Section \ref{sec:galileids}, one can still realize  this scenario using symmetries that do not commute with the Poincar\'e generators. Among various options the most minimal implementation requires only one Goldstone mode. We will dub such  system {\it type-I galileid}.

\item $\bar{P}_0 = P_0, \quad  \bar{P}_i =P_i, \quad \bar{J}_i = J_i+ \tilde{Q}_i$. \\
The commutation relations (\ref{unbroken algebra}) require 
\be \la{b1}
 [\tilde{Q}_i, \tilde{Q}_j ] = i \epsilon_{ijk} \tilde{Q}^k, 
\ee
which means that the $\tilde Q_i$'s are the generators of an internal $SO(3)$ group. In this scenario, we have at least six broken generators (the $K_i$'s and the $\tilde Q_i$'s) and six Goldstone bosons, since there are no inverse Higgs constraints one can impose. We will call {\it type-II framid} a condensed matter system described by this pattern of symmetry breaking. A possible order parameter consists of a triplet of vector fields $A_\mu^a$ that rotates under the internal $SO(3)$ symmetry and takes a vev $\langle A_\mu^a \rangle = \delta_\mu^a$, with $a = 1,2,3$.

%Once again, notice that any non-trivial extension of the internal $SO(3)$ group will necessarily lead to additional Goldstone modes. This is because such extensions are non-trivial only if (some of) the additional generators do not commute with the $\tilde Q_i$'s. Such non-commuting generators must be broken, or otherwise we would have some unbroken internal symmetries that do not commute with $\bar J_i$, and it would no longer be possible to interpret the latter as generators of some effective spatial rotations. Since these additional broken generators commute with all unbroken translations, the corresponding Goldsone modes cannot be eliminated by the inverse Higgs mechanism and belong to the spectrum of physical excitations.

\item $\bar{P}_0 = P_0 + Q, \quad  \bar{P}_i =P_i + Q_i, \quad \bar{J}_i = J_i$. \\
Once again, consistency with the commutation relations (\ref{unbroken algebra}) implies that the $Q^k$'s must transform like a 3-vector under rotations, as encoded in equation (\ref{3}). Therefore, this scenario can only be realized by adding symmetries that do not commute with the Poincar\'e group, like case 3 above.  The resulting pattern of symmetry breaking defines what we will call a {\it type-II galileid},  which we will elaborate on in Sect.~\ref{sec:galileids}.

\item $\bar{P}_0 = P_0 + Q, \quad  \bar{P}_i =P_i, \quad \bar{J}_i = J_i +\tilde{Q}_i$. \\
The generators $\tilde Q_i$ must again be the generators of an internal $SO(3)$, like in the scenario 4 discussed above. It then follows from the  algebra (\ref{unbroken algebra}) that these generators must commute with $Q$, which in the simplest implementation can be thought of as the generator of an internal $U(1)$ symmetry. We have therefore a total of seven broken generators ($Q$, $\tilde Q_i$, $K_i$), but because $[\bar P_i, K_j] = - i \delta_{ij} (\bar P_0 - Q)$, the boost Goldstones can be eliminated via inverse Higgs constraints. Thus, we expect four independent Goldstone modes. This symmetry breaking pattern defines a {\it type-II superfluid} and, in the non-relativistic limit, is realized in nature by the B-phase of superfluid He3~\cite{vollhardt2013superfluid}. In that case the generators $\tilde{Q}_i$ describe {\it spin}, which in a non-relativistic system with negligible spin-orbit couplings can be thought of as an internal $SO(3)$ symmetry. 

Relativistic type-II superfluids have recently been discussed in~\cite{Endlich:2013vfa} with particular emphasis on the peculiarities of their UV completion. In this respect, it is interesting to notice that there is no order parameter realizing their symmetry breaking pattern for which the inverse Higgs constraints correspond to removing gauge-redundant Goldstone fields \cite{Nicolis:2013sga}. For instance, a fairly minimal order parameter is an $SO(3)$ triplet of complex four-vectors with vev
\footnote{A slightly more minimal possibility would be a complex scalar plus a triplet of real four-vectors, with vevs
\be
\langle \Phi \rangle = e^{it} \; , \qquad \langle A^a_\mu \rangle = \delta^a_\mu \; .
\ee}
\be
\langle A^a_\mu(x) \rangle = e^{i t} \, \delta^a_\mu \; ,
\ee
which, when acted upon by the broken generators, yields seven independent Goldstone fields---three more than the necessary four.
It turns out that for all weakly coupled realizations of this symmetry breaking pattern, the inverse Higgs constraints always correspond to integrating out three {\it gapped} Goldstone fields from the Lagrangian \cite{Endlich:2013vfa}.

Like in the case of type-I superfluids, any non-trivial compact extension of the  $SO(3) \times U(1)$ group will inevitably  lead to additional Goldstone modes. Finally, notice that, like in the case of type-I superfluids, one could choose to promote the $U(1)$ internal symmetry to the monotonic internal diffeormophisms \eqref{khronon} without introducing additional Goldstones.

\item $\bar{P}_0 = P_0, \quad  \bar{P}_i =P_i + Q_i, \quad \bar{J}_i = J_i + \tilde{Q}_i$. \\
By the commutation relations (\ref{unbroken algebra}) and by the request that $Q_i$ and $\tilde Q_i$ commute with Poincar\`e one must have 
\ba \la{7}
 [ \tilde{Q}_i, \tilde{Q}_j] = i \epsilon_{ijk} \tilde{Q}^k, \qquad\quad   [ Q_i, Q_j]  = 0, \qquad\quad  [ \tilde{Q}_i, Q_j] = i \epsilon_{ijk} Q^k \; . 
\ea
This is the algebra of the three-dimensional Euclidean group $ISO(3)$. That is, the $\tilde{Q}_i$'s generate an internal $SO(3)$ symmetry, and the $Q_i$'s generate three-dimensional internal translations. In this scenario, we have a total of nine broken generators ($Q_i$, $\tilde{Q}_i$ and $K_i$), but we can have as few as three Goldstone excitations. This is because the commutation relations $[\bar P_0, K_i] = - i (\bar P_i - Q_i)$ and $[ \bar P_i, \tilde Q_j] = i \epsilon_{ijk} Q^k$ allows one to impose six inverse Higgs constraints to express the $K_i$ and  $\tilde{Q}_i$ Goldstones in terms of derivatives of the $Q_i$ ones~\cite{Nicolis:2013lma}. The minimal implementation in which the internal symmetry group is just $ISO(3)$ describes ordinary (isotropic) {\it solids}.
To see this at the level of the low-energy EFT, it is convenient to characterize this system in terms of an internal $SO(3)$ triplet of scalar fields $\phi^a$, which can be interpreted as the comoving coordinates of a solid's volume elements \cite{Dubovsky:2005xd, Son:2005ak, Nicolis:2013lma}, and shift under the internal $Q_i$'s, $ \phi^a \to \phi^a + c^a$. The expectation values
\be
\langle \phi^a (x) \rangle = x^a
\ee
realize the correct symmetry breaking pattern, and the three fluctuation fields $\pi^a$ defined by $\phi^a = x^a + \pi^a$ describe the three (acoustic) phonons of the solid, which are the only physical Goldstones that survive upon imposing all available inverse Higgs constraints for this breaking pattern.

Similarly to the superfluid case, also in this scenario it is possible to enlarge the internal symmetry group without increasing the number of Goldstone modes. In fact, one can even add an {\it infinite} number of internal generators and promote $ISO(3)$ to the group of internal diffeomorphisms with unit determinant, $\mbox{\it Diff}\,'(3)$. In terms of the triplet of scalars defined above,
\be \label{diff'}
\phi^a \to \xi^a(\phi) \; , \qquad \det \frac{\partial \xi^a}{\partial \phi^b} = 1 \; .
\ee
An infinite number of inverse Higgs constraints ensures that the number of Goldstones remains the same. Such a large internal symmetry group provides a low-energy effective description of ordinary {\it fluids}~\cite{Jackiw:2004nm, Dubovsky:2005xd}.  Notice that the number of independent Goldstone fields is still three, but only the longitudinal one features wave solutions---the fluid's sound waves. The two transverse Goldstones have a degenerate $\omega = 0 $ dispersion law, and can be thought of as the linearized progenitors of vortices.

There is an interesting intermediate case still featuring three Goldstones, where the internal group is the three dimensional {\it special affine} group, which is  finite dimensional but non-compact, and contains $ISO(3)$ as a subgroup. Its action on our triplet of scalars is 
\be
\phi^a \to c^a + M^a {}_b \, \phi^b \; , \qquad \det M =1 \;, 
\ee
where, unlike for the $\mbox{\it Diff}\,'(3)$ case, $c^a$ and $M^a {}_b$ are constant. Curiously, for this system the full $\mbox{\it Diff}\,'(3)$ is recovered as an accidental symmetry to lowest order in the derivative expansion \cite{Nicolis:2013lma}. In other words, at low enough energies such a system is indistinguishable from an ordinary fluid.

\item $\bar{P}_0 = P_0 + Q, \quad  \bar{P}_i =P_i + Q_i, \quad \bar{J}_i = J_i+ \tilde{Q}_i$. \\
Starting from (\ref{unbroken algebra}), it is easy to show that the commutation relations (\ref{7}) must still hold, and that $Q$ must commute with all other generators. This is the algebra of the three-dimensional Euclidean group (internal translations and rotations), supplemented by an extra $U(1)$ symmetry generated by $Q$.
We have a total of ten broken generators ($Q$, $Q_i$, $\tilde{Q}_i$, and $K_i$), but by imposing the same inverse Higgs constraints as in the previous scenario we are left with only four Goldstone bosons. If the internal symmetry group is exactly $ISO(3) \times U(1)$, we recover the pattern of symmetry breaking associated with {\it supersolids}~\cite{Son:2005ak}. A useful parameterization of the Goldstone excitations involves the same $\phi^a$ triplet of scalars we defined above for solids and fluids (case 7), as well as the superfluid phase field $\psi$ we defined for type-I superfluids (case 2), with symmetry breaking expectation values
\be
\langle \phi^a (x) \rangle = x^a \; , \qquad \langle \psi (x) \rangle = t \; .
\ee
The four independent fluctuation modes about these backgrounds describe the Goldstone excitations.

Once again, the internal symmetry group can be made infinite-dimensional without the need for additional Goldstone modes. By promoting again $ISO(3)$ to $\mbox{\it Diff}\,'(3)$ (eq.~\eqref{diff'}), one recovers the long-distance dynamics of a {\it finite-temperature superfluid} \cite{Nicolis:2011cs}. In this case two of the four Goldstones---the transverse ones---acquire a degenerate $\omega=0$ dispersion law, and can be identified with (linearized) vortex degrees of freedom in the normal fluid component. The remaining two Goldstones describe first and second sound.
If one further promotes the $U(1)$ constant shifts on $\psi$ to ``chemical shifts"~\cite{Dubovsky:2011sj}, 
\be
\psi \to \psi + f(\phi^a) \; , 
\ee
one obtains a {\it charge-carrying ordinary fluid}. This makes another Goldstone mode become degenerate, with $\omega = 0$, thus leaving one with ordinary sound waves only.
Alternatively, one could also choose to promote the $U(1)$ to a khronon-like symmetry \eqref{khronon}, although we are presently not aware of any medium that enjoys these symmetries. 

\end{enumerate}

\subsection{Summary}

In summary, we have identified eight possible condensed matter scenarios that can be produced in a Poincar\'e invariant theory. Six of them can be realised using additional {\it internal} symmetries, while for two of them (cases 3 and 5 above) we have to resort to extra {\it spacetime} symmetries that do not commute with the Poincar\'e group. 
%Here and in the rest of the paper we make the minimal choice of invoking extra spacetime symmetries only in the latter two cases, {\it i.e.},  when strictly needed. 

The results of our analysis are summarized in Table \ref{t1}, where for each scenario we display the minimum number of Goldstone modes and the corresponding  symmetry group. 
\begin{table}[t!]
\begin{center}
\begin{tabular}{ | l | c | c | c | c | c | c |}
\hline
\multirow{2}{*}{\bf \ \ \  System} & \multicolumn{3}{| c |}{\!Modified generators\!} & \multirow{2}{*}{\!\# G.B.\!} &{Internal} & {\!Extra spacetime\!} \\
%\hline
\cline{2-4} & \ \ $P_t$ \ \ & \ \ $P_i$ \ \ & $J_i$ & & { symmetries} & { symmetries} \\
\hline
1.  type-I framid &  &  &  & 3  & & \\
\hline
2.  type-I superfluid & \checkmark &  &  & 1 & $ U(1)$ & \\
\hline
3.  type-I galileid &  &  \checkmark &  & 1  &  & Gal \!(3+1,1) \tablefootnote{See for instance~\cite{Goon:2012dy} for a general definition of the groups Gal \!($d$ + 1, $n$).\label{note1}} \\
\hline
4.  type-II framid & &  & \checkmark & 6 & $SO(3)$ & \\
\hline
5.  type-II galileid &  \checkmark &  \checkmark & & 1  &   &Gal \!(3+1,1) \footref{note1} \\
\hline
6.  type-II superfluid\! & \checkmark &  & \checkmark & 4 & $SO(3) \times U(1)$ & \\
\hline
7.  solid & & \checkmark & \checkmark & 3 & $ISO(3)$ & \\
\hline
8.  supersolid &\checkmark  &\checkmark  & \checkmark & 4 & $\!ISO(3) \times U(1)\!$ & \\
\hline
\end{tabular}
\end{center}
\caption{\it \small The eight possible patterns of symmetry breaking discussed in the text. The checkmarks denote whether a translation or rotation is mixed with another symmetry. 
%and weather a symmetry breaking pattern requires extra spacetime symmetries to be realised. 
The ``$\# \ G.B.$" column displays the {\it minimum} number of Goldstone modes necessary to non-linearly realize the broken symmetries, whereas the last two columns display the (minimal) symmetry group needed to realize the desired breaking pattern.
}
\label{t1}
\end{table}
Four of these scenarios---type-I and type-II superfluids, solids (with ordinary fluids being a special case), and supersolids---are already known to be realized in nature.\footnote{There is a controversy as to whether supersolids have actually been observed. However, systems that qualify as supersolids according to our low-energy EFT criteria---symmetries of the gapless excitations' dynamics---clearly exist. Take for instance a very porous but fairly rigid material, like a metal (open-cell) foam, and immerse it in superfluid helium, which will then fill all the voids of the material. At distances much bigger than the typical cell size, the dynamics will be those of a superfluid coupled to the vibrational modes of a solid---a supersolid. (A  more expensive example is that of a huge empty building with internal walls and rooms and hallways and staircases---but no closed doors---filled with superfluid helium\dots) 
}
%%
%%---at least in a purely non-relativistic setting. 
We refer the reader to the literature for the details of their effective field theories, which have been studied extensively in recent years \cite{Son:2002zn,Son:2005ak,Dubovsky:2005xd,Dubovsky:2011sj,Nicolis:2013lma, Endlich:2012pz}.
The next two sections are devoted to a discussion of type I and type II framids---the only scenarios that admit {\it gapless} Goldstone modes associated with the spontaneously broken boosts.  Finally, in Section \ref{sec:galileids} we examine the two remaining scenarios, 3 and 5.

\section{Framids}
\label{sec-3}

For simplicity, we will first develop some intuition by focusing on the simpler type-I case, and we will show that framids are not just a simpler description of more mundane states of matter such as solids. We will then derive the low-energy effective action for both type-I and type-II framids using the coset construction~\cite{Callan:1969sn,Coleman:1969sm,Volkov:1973vd,ogievetsky:1974ab}. This approach will clearly show how type-I framids are in essence just a special case of type-II framids. %Finally, we will conclude by providing some speculations as to why framids have not been observed in nature so far. 

\subsection{Type-I framids}\label{type-I}

As   discussed above, a possible order parameter for type-I framids is a vector local operator $A_\mu(x)$, acquiring an expectation value
\be
\langle A_\mu (x) \rangle = \delta_\mu^0
\ee
on the ground state. This expectation value breaks boosts, and the three corresponding Goldstone modes $\vec \eta (x)$---the {\it framons}---can be introduced by parametrizing the fluctuations of $A_\mu(x)$ as
\be \la{Aparam}
A_\mu(x) = \big(e^{i \vec \eta(x) \cdot \vec K}  \big)_\mu {}^\alpha \, \langle   A_\alpha (x)  \rangle \; ,
\ee
where the $\vec K$'s are the boost generators in the relevant representation (the four-vector one).
This parametrization is particularly convenient because the vector field $A_\mu (x)$ turns out to have a constant norm, i.e.~$A_\mu A^\mu = -1$. Thus the low-energy effective Lagrangian for the Goldstones $\vec \eta$ can be obtained by writing the most general Poincar\'e-invariant Lagrangian with at most two derivatives for a vector field with constrained norm. Up to total derivatives this is simply\footnote{The rationale behind our definition of the couplings $M_i^2$ will become clear in Section \ref{sec:coset}.}
\begin{equation} \label{lagrangianA}
\mathcal{L} = -\sfrac{1}{2} \Big\{ M^2_3 (\partial_\mu A^\mu)^2 + M^2_2 \, (\partial_\mu A_\nu)^2 + (M^2_2 - M^2_1) (A^\rho \partial_\rho A_\mu) ^2\Big\}.
\end{equation}
Although the kinetic terms do not involve just the usual gauge invariant combination $F_{\mu \nu} F^{\mu \nu}$, ghost instabilities are avoided because the norm of $A_\mu$ is constrained. Put another way, the vector field $A_\mu(x)$ is just a placeholder for the Lorentz-covariant combination of Goldstone fields \eqref{Aparam}, and so only three independent degrees of freedom enter the variational principle.
In fact, we will see in a moment that the three coefficients $M_i^2$ can always be chosen in such a way that all three Goldstones are well-behaved. 

The Lagrangian (\ref{lagrangianA}) is particularly simple because it econdes {\it infinitely many} interactions through a finite number of couplings: By substituting the expression \eqref{Aparam} in eq.~(\ref{lagrangianA}), one obtains a Lagrangian for the $\vec \eta \,$'s that includes interactions with an arbitrarily large number of Goldstones but just two derivatives. This should be contrasted  with the case of ordinary matter (solids, fluids, superfluids, supersolids), where each Goldstone field in the Lagrangian carries at least one derivative, and so at the two-derivative level there are no interactions \cite{Dubovsky:2005xd, Endlich:2012pz}. From this viewpoint, the framid Lagrangian is similar to the chiral Lagrangian for the QCD pions (more below).

In order to study the particle content, it proves useful to separate the vector $\vec \eta$ into its longitudinal and transverse components:
\begin{equation}
\vec \eta = \vec \eta_L + \vec \eta_T\, , \qquad\quad  \vec \partial \cdot \vec \eta_T = 0\, , \qquad  \vec \partial \times \vec \eta_L = 0 \; ,
\end{equation}
The quadratic Lagrangian then reads
\begin{align} \label{lag-quad}
{\cal L}_2 \ = \sfrac12 {M^2_1} \Big[\dot \vp^{\, 2} - c_L^2 \, (\vec{\partial} \cdot \vp_L)^2   - c_T^2 \, \partial_i \eta_T^j \, \partial_i \eta_T^j \Big]\, ,
\end{align}
where we have introduced the transverse and longitudinal propagation speeds, 
\be \label{speeds}
c_T^2 = \frac{M^2_2 }{M^2_1} \; , \qquad\qquad  c_L^2 = \frac{M^2_2+M^2_3}{M^2_1}\, .
\ee
From eq.~(\ref{lagrangianA}), it is clear that the parameter $M_1^2$ can always be factored out of the Lagrangian as a reference scale, and after doing that all interactions are completely determined by the two propagation speeds $c_T$ and $c_L$. For instance, the cubic and quartic interaction terms read
\begin{align}
{\cal L}_3 \ =& \ M^2_1 \left[\, (c_T^2 -1)\, \vp \cdot \partial \eta \cdot {\dot \vp} + (c_L^2 - c_T^2) \, [\partial \eta]\,  \vp \cdot {\dot \vp}\, \right] \, \label{L3},\\[3mm]
{\cal L}_4 \ =& \ \sfrac12 {M^2_1} \Big[ \left (\sfrac43 - c_T^2\right) \vp^{\, 2} \, \dot \vp^{\, 2} -  \left(\sfrac13 + c_L^2 - c_T^2\right) (\vp \cdot \dot \vp) ^2  +  (1 - c_T^2) \, \vp \cdot \d \eta \cdot \d \eta^T \cdot \vp  \label{L4} \\
&+ \sfrac13 {c_T^2} \, (\partial \eta \cdot \eta)^2  - \sfrac13 {c_T^2} \, [\d \eta ^{\, T} \d \eta] \, \vp^{\, 2} 
- \sfrac13 {(c_L^2 - c_T^2)} \, [\d \eta]^2 \, \vp^{\, 2} - \sfrac23 {(c_L^2 - c_T^2 )} \, [\d \eta] \, \vp \cdot \d \eta \cdot \vp \Big]\nonumber,
\end{align}
where $\partial \eta$ denotes the matrix with entries $(\partial \eta)_{ij} = \partial_i \eta_j$, $\partial \eta^T$ its transpose (and {\it not} its restriction to the transverse modes), and the brackets $[\dots]$  the trace of the matrix within.

A particularly interesting limit to consider is the ultra-relativistic one with $c_L, c_T \to 1$. In this case, the Lagrangian for the framons reduces to
\begin{equation} \label{almost-pions}
{\cal L}_{c_L , c_T \to 1}\ = \ -\sfrac12 {M_1^2} \left[(\partial_\mu \vec \eta \, )^2 - \sfrac13 \partial_\mu \eta^i \partial^\mu \eta^j\left(\eta_i \eta_j - \delta_{ij} \vec \eta^{\, 2}\right) + {\cal O}(\eta^6) \right] \, .
\end{equation}
In this limit we are in the presence of an {\it enhanced symmetry}, which follows from the fact that the Lagrangian for the order parameter $A^\mu$ reduces to the single term 
\be
{\cal L}_{c_L , c_T \to 1} \propto \partial_\mu A^\alpha \partial^\mu A_\alpha
\ee 
[see eqs.~\eqref{lagrangianA} and~\eqref{speeds}].  
Such a term enjoys {\it two} independent Lorentz symmetries, acting on the derivatives ($\mu$ index) and on the fields ($\alpha$ index) separately. Interestingly, a similar enhanced ``spin" symmetry applies to gauge-theory amplitudes in certain infinite-momentum limits~\cite{ArkaniHamed:2008yf}. 
One can thus view the Lorentz group acting on $A^\alpha$ as an {\it internal} group that is spontaneously broken down to internal rotations, i.e. $SO(3,1) \rightarrow SO(3)$. In particular, the boosts of this internal Lorentz group are non-linearly realized on the Goldstones. The  spacetime Lorentz group that acts on the derivatives instead remains  unbroken, because the expectation value of $A_\alpha$ is constant.  This is very clear from the form of~\eqref{almost-pions}, in which rotations are linearly realized on the $\vec \eta$ triplet and the indices carried by the derivatives are never contracted with the indices carried by the $\vec \eta \, $'s, i.e.~there are no ``spin-orbit'' couplings. Note that we are just a factor of $i$ away from the symmetry breaking pattern relevant for the QCD  pions, $SU(2) \times SU(2) \rightarrow SU(2)$. Indeed, the chiral Lagrangian for the pions $\pi^a$ at the two-derivative level can be obtained from~\eqref{almost-pions} upon the formal replacement $\eta^a = i  \pi^a$ (and $M_1^2 \to -f^2_\pi$), which has the effect of changing the relative sign between the quadratic and quartic operators. This relative sign is determined by the curvature of the coset manifold, which is positive for $SU(2) \times SU(2) / SU(2)$ and negative for $SO(3,1) / SO(3)$.

Another interesting limit is that with small $c_L$ and $c_T$. In this case the strong coupling scale of the theory will not be simply $M_1$. In order to estimate it,  we need to keep separate scaling dimensions for energies $\omega$ and momenta $k$. For $c_T \ll 1$ and $c_L \ll 1$, the canonically normalized framon field $\vp_c = M_1 \vp $ has scaling dimensions $[k]^{3/2} [\omega]^{-1/2}$, as apparent by inspection of the quadratic action. 
When writing interactions in terms of the canonically normalized fields, the scale $M_1$ appears at the denominator with appropriate powers. Therefore, the terms with the {\it lowest} strong coupling scale will be those in which $c_T$ and $c_L$ appear as small corrections to order one coefficients. For instance, the first term in the cubic Lagrangian~\eqref{L3} violates unitarity at smaller energies than the second term, in the limit of small propagation speeds. Similarly, terms with a higher number of spatial derivatives constrain the strong coupling scale more tightly, because $\partial_t \sim c_{L,T} \cdot \partial_i \ll \partial_i$. In summary, we need to inspect the two (cubic and quartic) operators 
\begin{equation} \label{small c}
{\cal S}_{int} \ = \ \int dt \, d^3x \, (1 - c_T^2) \Big[- \frac{1}{M_1} \vp_c \cdot \partial \eta_c \cdot {\dot \vp_c} +  \frac{1}{2 M_1^2}  (\eta_c \cdot \d \eta_c)^2 \Big]\, .
\end{equation}
Since $\vec \eta_c$ has  dimensions $[k]^{3/2} [\omega]^{-1/2}$, it is easy to show that the dimensions of $M_1$ are $[k]^{5/2} [\omega]^{-3/2}$. Therefore,  if we assume that the two sound speeds are comparable, $c_L \sim c_T \ll 1$, we conclude purely on dimensional grounds that the energy and momentum strong coupling scales must be
\begin{equation} \label{strong-coupling}
E_{\rm strong} \sim \ M_1 \, c_{L,T}^{\, 5/2}\, , \qquad p_{\rm strong} \sim \ M_1 \, c_{L,T}^{\, 3/2} \; .
\end{equation}

%%%%%%%%%%%%%%%%%%%%%%%%%%%%%%%%%%%%%%%%%%%%
%%%%%%%%%%%%%%%%%%%%%%%%%%%%%%%%%%%%%%%%%%%%
\subsection{Is the framid a solid in disguise?} \label{sec_amplitude}

Type-I framids and solids have the same number of Goldstones, which in both cases form a triplet  under the unbroken rotations.  An apparent feature of the framons $\vec \eta$ is the presence of interaction terms (eqs.~\eqref{L3}-\eqref{L4}) with two derivatives, whereas the phonons in a solid always carry at least one derivative per field, and so a cubic interaction has three derivatives, a quartic interaction has four, and so on \cite{Dubovsky:2005xd, Endlich:2012pz}.  This should be enough to conclude that framids and solids are physically inequivalent systems.  However, one might still suspect that some field redefinition could turn framons into phonons. 

In order to make sure that this is not the case, we have calculated the $2\to2$ tree-level scattering amplitude for framons. The naive expectation based on the simple derivative counting is correct: amplitudes that would scale as $E^4$ for the phonons in a solid, scale indeed as $E^2$ in the case of framons, and no magical cancellations happen. This is also obvious from the previous section where we saw we that can tune the framid into  the relativistic invariant $SO(3,1) / SO(3)$ $\sigma$-model, which is well known to have amplitudes that scale like $E^2$ at low-energy.

For example,  the elastic scattering amplitude for the head-on collision of two longitudinal framons of equal energy $E$ is 
\be \label{amplitude}
i {\cal M}_{LL\rightarrow LL} = - 2i \, \frac{E^2}{c_T^2 M_1^2} \times  f(\theta)
\ee
where $f(\theta)$ is an order-one function of the scattering angle,
\be
f(\theta) = \sfrac{(1+c_L^2)^2 + (c_L^4 - 6 c_L^2 - 3) \cos^2\theta + 4 \cos^4 \theta  - 2 (c_T/c_L)^2 (1- \cos^2 \theta) (c_L^4 +(1- 2 c_L^2) \cos^2\theta  ) }{ 1- \cos^2 \theta}\, .
\ee
As evident from the above expression, amplitudes really do scale as $E^2$ at low energies. It is not possible to make~\eqref{amplitude} vanish for all scattering angles with a specific choice of $c_L$ and $c_T$, so there is no tuning of the Lagrangian coefficients that can  turn a framid into a solid. We also note  that, while comparing~\eqref{amplitude} with the general estimate of the strong coupling scale~\eqref{strong-coupling}, one should keep in mind that the $2\rightarrow2$ scattering amplitude scales as velocity to the third power, $[{\cal M}] = [E/p]^3$ (see e.g.~\cite{Endlich:2010hf}). This means that~\eqref{amplitude} reaches the unitary bound when ${\cal M} \sim c^3$,  {\it i.e.}, at $p_{\rm strong} \sim M_1 c^{3/2}$, $E_{\rm strong} \sim M_1 c^{5/2}$, as correctly estimated in~\eqref{strong-coupling}.

\subsection{Coset Construction: type-II $\to$ type-I}
\la{sec:coset}

Having built some general intuition about framids, we are now ready for a more systematic analysis. As discussed in Section \ref{SSB}, type-I and type-II framids are characterized by very different order parameters: a single vector field with a time-like vev for the former,  a triplet of vectors with mutually orthogonal space-like vevs for the latter. This striking difference hides the fact that the low-energy effective action of type-I framids is just a particular limit of that of the type-II ones. However, this becomes immediately apparent when such effective actions are derived using the coset construction~\cite{Callan:1969sn,Coleman:1969sm,Volkov:1973vd,ogievetsky:1974ab}. In what follows we will resort to this technique, and while we will try to be as self-contained as possible, the most ``coset-phobic'' readers are referred to~\cite{Delacretaz:2014oxa} for a general but concise review of the coset construction ideology. The most impatient ones can instead skip directly to the final results: the low-energy effective Lagrangians (\ref{effaction}) and (\ref{actionI}), with the relevant quantities  defined in eqs.~(\ref{covdev}). Finally, the simply uninterested ones can safely skip to sect.~\ref{why not} without loss of continuity.

As discussed in Section \ref{SSB}, type-II framids are characterized by an internal broken $SO(3)$ symmetry, which combines with the broken spatial rotations yielding an unbroken diagonal $SO(3)$. In summary, the symmetry breaking pattern reads 
\ba \la{pattern}
\begin{array}{lcl}
\mbox{unbroken} &=&  \l\{
\begin{array}{ll}
\bar{P}_t \equiv P_t&  \quad\qquad\qquad \,\,\mbox{time translations} \\
\bar{P}_i \equiv P_i &  \quad\qquad\qquad \,\,\mbox{spatial translations}  \\
\bar{J}_i \equiv J_i + S_i &  \quad\qquad\qquad \,\,\mbox{rotations}
\end{array}
\r. \nonumber 
\\ \\
\mbox{broken} &=&  \l\{
\begin{array}{ll}
S_i & \qquad\qquad\qquad\qquad\quad\,\,\,   \mbox{internal $SO(3)$} \\
K_i &  \qquad\qquad\qquad\qquad\quad\,\,\,   \mbox{boosts} 
\end{array}
\r. \nonumber 
\end{array}
\ea
The starting point of the coset construction is the coset parametrization
\be \la{cospar}
\Omega (x) = e^{i x^\mu \p_\mu} e^{i \eta^j (x)K_j} e^{i \theta^j (x)S_j}  \, ,
\ee
which is nothing but a parametrization of the most general symmetry transformation that is non-linearly realized. As such, it contains the generators of the spontaneously broken symmetries ($K_i$ and $S_i$) together with their respective Goldstones ($\eta_i$ and $\theta_i$), but also the generators of unbroken translations, which are always non-linearly realized on the coordinates $x^\mu$. The transformation properties of the coordinates and the Goldstone fields under a generic element $g$ of the symmetry group is determined by the equation:
\be \la{trancos}
g \, \Omega (x, \eta, \theta) = \Omega (x', \eta', \theta') \, h (\eta, \theta, g),
\ee
where $h (\eta, \theta, g)$ is an element of the unbroken subgroup that in general depends on the Goldstones as well as $g$. For any given $g$, the explicit form of $\Omega (x', \eta', \theta')$  and $h (\eta, \theta, g)$ can be calculated explicitly by moving  $g$ past $\Omega (x, \eta, \theta)$ using the algebra of the Poincar\'e and internal $SO(3)$ groups and casting the end result as a product of a non-linearly realized symmetry transformation and an unbroken one.

In order to construct an effective action that is invariant under broken and unbroken symmetries alike, one needs to calculate the Maurer-Cartan one-form $\Omega^{-1} \d_\mu \Omega$  and expand its coefficients in the basis of generators $\l\{\bar P_\mu, K_i, S_i,\bar J_i \r\}$. Once again, such a calculation can be carried out solely using the symmetry algebra, and the final result can be cast in the following form:
\be \la{MCform}
\Omega^{-1} \d_\mu \Omega \, = i e_\mu{}^\nu \l( \bar P_\nu + D_\nu \eta^i K_i + D_\nu \theta^i S_i + {\cal A}_\nu^i \bar J_i \r).
\ee
With some hindsight, we have denoted the coefficients of such an expansion in a suggestive way.
In fact, the transformation properties of such coefficients follow directly from (\ref{trancos}) and are such that 
\begin{itemize}
\item $e_\mu{}^\nu$ plays the role of a vierbein, in the sense that it can be used to build a volume element $d^4 x \det (e)$ that is invariant under all the symmetries.
\item $D_\nu \eta^i$ and $D_\nu \theta^i$ should be thought of as covariant derivatives of the Goldstone fields. They are non-linear in the Goldstones and transform linearly under the unbroken symmetries. Any contraction of $D_\nu \eta^i$ and $D_\nu \theta^i$ that is invariant under the unbroken symmetries (in this case, the rotations generated by $\bar{J}_i$), yields a quantity that is actually invariant under all the symmetries. For compactness we are using a Lorentz covariant-looking notation, but the $\nu=0$ and $\nu=i$ of such covariant derivatives have to be treated independently. 
\item ${\cal A}_\nu^i$ acts as a connection, that can be used to define higher covariant derivatives of the Goldstone fields, as well as covariant derivatives of additional matter fields. In what follows we will not need this connection, because we will focus on the Goldstone sector and work at lowest order in the derivative expansion.
\end{itemize}
We should stress that the explicit form of $e_\mu{}^\nu, D_\nu \eta^i, D_\nu \theta^i$ and ${\cal A}_\nu^i$ crucially depends on how the coset $\Omega(x)$ is parametrized. Different parametrizations are related by a redefinition of the Goldstone fields. The parametrization that we have chosen in (\ref{cospar}) is particularly convenient because in this case $e_\mu{}^\nu$ is just a Lorentz boost $\Lambda_\mu{}^\nu$ with rapidity $\vec \eta$, and thus its determinant is trivial, $\det (e) =1$.

An explicit calculation of the Maurer-Cartan form yields the following covariant derivatives:
\begin{subequations} \la{covdev}
\ba
D_\mu \eta_i &=& (\Lambda^{-1})_\mu{}^\nu \d_\nu \eta^j \l\{ \delta_{ji} + \l[\fr{\eta - \sinh \eta}{\eta^3} \r] ( \eta_j \eta_i - \delta_{ji} \vec{\eta}^{\, 2} )  \r\} \la{covdeveta} \\
D_\mu \theta_i &=& (\Lambda^{-1})_\mu{}^\nu \d_\nu \theta^j \left\{\delta_{ji} + \l[\frac{1 - \cos \theta}{\theta^2}\r] \theta ^k \epsilon_{kji} + \l[\frac{\theta - \sin \theta}{\theta^3} \r](\theta_j \theta_i - \theta^2 \delta_{ji})\right\},\quad 
\ea
\end{subequations}
where $\Lambda^{-1} = \exp(- i \vec \eta \cdot \vec K), \eta \equiv \sqrt{\vec \eta \, ^2}$ and $\theta\equiv \sqrt{\vec \theta \, ^2}$ . Notice that an expansion in powers of $\eta$ and $\theta$ yields only even powers, and therefore $D_\mu \eta_i$ and $D_\mu \theta_i$ are analytic in $\vec \eta$ and~$\vec \theta$. These covariant derivatives are the main building blocks one should use to write down the low-energy effective action. At the 2-derivative level this is:
\ba
\mathcal{L}_{\rm type \, II} &=& \sfrac{1}{2} \Big\{ M^2_1 (D_0 \vec \eta \, )^2 - M^2_2 \, (D_i \eta_j)^2  - M^2_3 (D_i \eta^i)^2   \la{effaction} \\
&& \qquad  + M^2_4 \, D_i \theta^i + M^2_5 (D_0 \vec \theta \, )^2 - M^2_6 \, (D_i \theta_j)^2  - M^2_7 (D_i \theta^i)^2 - M^2_8 \, D_i \theta_j D^j \theta^i \nonumber \\
&& \qquad\quad + 2 M^2_9 \,  D_0 \theta_i D_0 \eta^i -2  M^2_{10} \, D_i \theta_j D^i \eta^j - 2 M^2_{11} D_i \theta^i D_j \eta^j - 2 M^2_{12} \, D_i \theta_j D^j \eta^i    \Big\}. \nonumber
\ea

We could have also derived this action in the same way as we did above for type-I framids, that is, by introducing the Goldstones fields directly at the level of a specific order parameter (which transforms linearly under all the symmetries), 
\be
A_\mu^a(x) = \big(e^{i \vec \eta(x) \cdot \vec K}  \big)_\mu {}^\alpha \, \big(e^{i \vec \theta(x) \cdot \vec S}  \big)^a {}_b \, \langle   A^b_\alpha (x) \rangle \; , \qquad \qquad \langle   A^b_\alpha (x) \rangle = \delta^b_\alpha \; , 
\ee
and  then writing down the most general action for that order parameter. Already at the two-derivative level there are many invariants, and in this language it less straightforward to know when all possibilities have been exhausted. This is because one  in principle can have several factors of undifferentiated $A^a_\mu$'s, but upon contracting the indices there can be dramatic simplifications. For instance, of the two partial contractions
\be
B^{ab} (x) \equiv A_\mu^a A^{\mu \, b} \; , \qquad C_{\mu\nu}(x) \equiv A_\mu^a A_{\nu}^a \; , 
\ee
the first is trivial, $B^{ab} = \delta^{ab}$, while the second is not (it is the projector onto the 3D space locally spanned by the $A^a_{\mu}$'s.) In the Goldstone language instead, the Lagrangian terms cannot be ``dressed" by undifferentiated Goldstones, because the Goldstone fields do not transform covariantly. On the other hand, the equivalence of certain terms up to total derivatives can be obvious in the order parameter language, like for instance for
\be
\d_\mu A^a_\nu \,  \d^\nu A^{\mu \, a } \quad \leftrightarrow \quad (\d^\mu A^a_\mu)^2 \; ,
\ee
while being totally obscure  in terms of the Goldstones  (more below).  Notice finally that here too there is an enhanced symmetry case. If we restrict the Lagrangian to terms in which derivatives are never contracted with fields,
\be
{\cal L}_{\rm enhanced} = \sfrac12 \Big[ \tilde M_1^2 \, \partial_\mu A^a_\alpha \, \partial^\mu A^{\alpha \, a} 
+ \tilde M_2^2 \, A^{\alpha \, a} A^{\beta \, a} \, \partial_\mu A^b_\alpha \, \partial^\mu A^b_\beta 
\Big] \; ,
\ee
we are guaranteed to get Goldstone dynamics that respect a non-linearly realized {\it internal} Lorentz symmetry ($\alpha$, $\beta$ indices) as a well a linearly realized spacetime Poincar\'e symmetry ($\mu$ indices). In this case, spacetime symmetries are unbroken and the spontaneous breaking pattern for internal symmetries is
\be
SO(3,1) \times SO(3) \to SO(3)  \; .
\ee
Upon imaginary redefinitions of the $\vec \eta$ Goldstone fields, our action is formally equivalent to the chiral Lagrangian for the coset
\be
 SU(2) \times SU(2) \times SU(2) \, / \, SU(2)   \; .
\ee

The effective Lagrangian for type I-framids---which features no internal $SO(3)$ and no broken rotations---can be obtained from the action (\ref{effaction}) simply by setting to zero all the rotation Goldstones $\theta_j$. 
The reason is purely formal, and we don't see any physical reason why that should be the case \cite{Nicolis:2013lma}: postulating that rotations are broken and then ignoring the corresponding Goldstones is not physically equivalent to saying that rotations are unbroken.
However, since all our results follow from the coset parametrization \eqref{cospar}, it is clear that setting $\theta_j = 0$ there is equivalent to never introducing the internal $SO(3)$ and the associated spontaneous breaking in the first place. In other words, eq.~\eqref{cospar} with $\theta_j = 0$ is the correct coset parametrization for type-I framids. 
%Clearly the Goldstones $\theta_j$ should not appear in the action of a type I framid simply because such system does not feature any  broken internal $SO(3)$ symmetry. The crucial point, though, is that the transformation properties of the remaining covariant derivatives  are solely determined by the commutation relations between the $K_i$'s and all the other generators. Since $[K_i, \bar J_j] = [ K_i, J_j]$, the $D_\mu \eta_i$'s transform the same way under rotations generated by $\bar J_i$ and by $J_i$, which means that the contractions appearing in the first line of (\ref{effaction}) are invariant quantities even in the case of type I framids. 
We thus get
\be \la{actionI}
\mathcal{L}_{\rm type \, I} = \sfrac{1}{2} \Big\{ M^2_1 (D_0 \vec \eta \, )^2 - M^2_2 \, D_i \eta_j D^i \eta^j - M^2_3 (D_i \eta^i)^2 \Big\}.
\ee
The careful reader may have noticed that this Lagrangian does not include a  $D_i \eta_j D^j \eta^i$ term. This is because such a contraction is equivalent to $(D_i \eta^i)^2$ up to an integration by parts, even though this cannot be immediately deduced just by looking at the covariant derivatives (\ref{covdeveta}). It is easier to prove this by working at the level of the order parameter $A_\mu$. Gapless fluctuations around its time-like vev can be parametrized using a boost matrix $\Lambda = e^{i \vec \eta \cdot \vec K}$, as in eq.~\eqref{Aparam}. Then, starting from (\ref{MCform}) it is easy to show that derivatives of the order parameter are related to the covariant derivative $D_\mu \eta^i$ introduced above by
\be \la{DetatoDA}
D_\mu \eta^i = (\Lambda^{-1})_\mu{}^\nu (\Lambda^{-1})^{i \rho} \, \d_\nu A_\rho.
\ee
Using the properties of Lorentz matrices as well as the fact that $A_\mu A^\mu = -1$, it follows from (\ref{DetatoDA}) that 
\be
(D_i \eta^i)^2 = (\d_\mu A^\mu)^2\; , \qquad D_i \eta_j D^j \eta^i = \d_\mu A_\nu \,  \d^\nu A^\mu \; , 
\ee
and this proves that these two terms are equivalent up to integrations by parts.\footnote{Incidentally, this is no longer true on a curved space-time, which is why Einstein-aether theories of gravity admit four independent parameters at the two-derivative level \cite{Jacobson:2000xp}.} (Analogous considerations apply to the possible single covariant derivative term $D_i \eta^i$, which turns out to be a total derivative:
$ D_i \eta^i = \d_\mu A^\mu $.)
Similarly, it is easy to show that 
\be
(D_0 \vec \eta \, )^2 =  (A^\rho \partial_\rho A_\mu) ^2 \; , \qquad (D_i \eta_j)^2  -(D_0 \vec \eta \, )^2 = (\partial_\mu A_\nu)^2 \; ,
\ee
which proves that the couplings $M_i^2$ that appear in the Lagrangian (\ref{actionI}) are indeed the exact same couplings that appeared in eq.~(\ref{lagrangianA}).

%%%%%%%%%%%%%%%%%%%%%%%%%%%%%%%%%%%%%%%%%%%%
%%%%%%%%%%%%%%%%%%%%%%%%%%%%%%%%%%%%%%%%%%%%
\section{Why don't we see framids in nature?}\label{why not}

Framids do not seem to be realized in nature. According to the classification of Sec.~\ref{SSB}, they correspond to legitimate spontanuous breaking patterns of Lorentz symmetry, with framids of type-I realizing {\it the simplest} symmetry breaking pattern of all. 
Moreover, the low-energy EFT characterizing the dynamics of their Goldstone excitations seem to make perfect sense, with no sign of instabilities nor of any other obvious pathologies. Given nature's generosity when it comes to condensed matter systems, why doesn't it give us framids?

\subsection{Where is the stuff}

The first, intuitive guess is that condensed matter systems are necessarily made up of ``stuff", and there must exist collective excitations corresponding to locally {\it displacing}  this stuff. In EFT terms, this means that there must exist long-distance fields that serve the purpose of local {\it position} degrees of freedom, like the comoving coordinates $\phi^a(x)$ of solids and fluids (see sect.~\ref{SSB}). On the other hand, the framid's Goldstones $\vec \eta(x)$ only parametrize the local {velocity} of the medium, and are thus unsuited to describe the excitations of ``stuff''.

However appealing, this logic blatantly fails already for superfluids: there, despite there being some stuff, quantum effects in the form of Bose statistics and Bose-Einstein condensation are such that standard position degrees of freedom are absent from the low-energy EFT description. Rather, the low-energy excitations are parameterized by the fluctuations of a single (scalar) field $\psi$ taking an expectation value {\it in time}, $\langle \psi \rangle = \mu t$ (see sect.~\ref{SSB}). So, the only positional degrees of freedom we can talk about for a superfluid concern temporal ``positions". 
Clearly, the ``stuff'' intuition is not of much help when it comes to condensed matter systems with important quantum effects.

A more refined  argument is attempted in the following subsection. It does not work either, but looking at the reasons of its failure allows us to draw some interesting and general  conclusions on non-relativistic EFTs.

\subsection{Small velocities\dots}

All condensed matter systems that we create in the lab are extremely non-relativistic, in the sense that their stresses, pressures, and internal energy densities are much smaller than their mass densities (in natural units), the propagation speeds of their excitations are extremely sub-luminal, etc.

From the microscopic viewpoint, we understand why this is the case: condensed matter in the lab is made up of atoms, whose typical binding and interaction energies are much smaller than
the atomic mass
\footnote{Ultimately, this is due to the weakness of electromagnetic interactions, $\alpha \ll 1$, and to the smallness of the electron-to-nucleon mass ratio, $m_e/m_p \ll 1$. For ordinary solids, $p/\rho$ is roughly controlled by the ratio between 
atomic binding energy and mass, which is of order $\alpha^2 m_e/m_p\sim 10^{-7}$.}.
Putting together atoms at low temperatures, we can only create non-relativistic materials.

However, from the low-energy EFT viewpoint, the property of being highly non-relativistic should be {\it technically natural}: an unconventional physicist  completely ignorant about the constituents of matter but well versed in EFT techniques,
should not need to invoke the microscopic argument we just gave to explain why it is perfectly ``natural" to have non-relativistic substances. Given how many such substances there are in nature, if the naturalness argument were to fail here, we see no reason why we should keep applying it to particle physics.
So, one possibility for why we do not see framids in the lab would be  that it is not technically natural for them to have small sound speeds.

Let us show that this argument does not work, by looking at  type-I framids for simplicity. 
As we remarked, their low-energy effective action is completely determined by  three parameters only: the transverse and longitudinal Goldstone speeds $c_T$ and $c_L$, and the overall scale $M_1$. As pointed out in sect.~\ref{type-I}, the interactions become particularly simple in the limit of small propagation speeds---the only cubic and quartic interactions that survive are shown in eq.~(\ref{small c}). We see that the strength of these interactions is of $\mathcal{O}(1)$ in units of $M_1$, which might suggest that the propagation speeds receive $\mathcal{O}(1)$ loop corrections. However, one should take into account that the UV cut-off of the loop integrals also 
depends on the propagation speeds---see eq. (\ref{strong-coupling}). 
Let us then consider the action for type-I framids in the non-relativistic limit,
\be \la{smallcS}
{S} \ \simeq M_1^2 \int d^3 x dt \l\{ \tfrac{1}{2} [\dot \vp^{\, 2} - c_L^2 \, (\vec{\partial} \cdot \vp_L)^2   - c_T^2 \, \partial_i \eta_T^j \partial_i \eta_T^j ] - \vp \cdot \partial \eta \cdot {\dot \vp} +  \tfrac{1}{2} (\vp \cdot \d \eta)^2 \r\}\, ,
\ee
where $\vec \eta_L$ and $\vec \eta_T$ are once again the longitudinal and transverse parts of $\vec \eta$. Radiative corrections to the propagation speeds can be derived by considering the 1PI vertex with two external legs, $\Gamma^{(2)}(E, \vec k)$, and isolating the part proportional to the square of the external 3-momentum $\vec k$. At one-loop, $\Gamma^{(2)}$ receives contributions from diagrams with two different topologies. If we assume for simplicity that the two propagation speeds are comparable, i.e.~$c_L \sim c_T \equiv c_s$ (the  ``speed of sound"), then we can use the strong coupling scales $E_{\rm strong}=  c_s \, p_{\rm strong} \sim c_s^{5/2} M_1$ derived in section \ref{type-I} to  estimate the correction to $c_s$:
%
%\begin{subequations}
\ba
\parbox[t][8.5mm][b]{20mm}{
\begin{fmfgraph}(65,25) 
\fmfleft{i1}
\fmfright{o1}
\fmfv{decor.shape=circle, decor.filled=full, decor.size=1.5thick}{v1,v2}
\fmf{plain,tension=3}{i1,v1}
\fmf{plain,tension=3}{v2,o1}
\fmf{plain,left}{v1,v2}
\fmf{plain,left}{v2,v1}
\end{fmfgraph}
} \quad\,\,\,\,\, &\sim& \int dE d^3 p \, (M_1^2 E p)^2 \l[\fr{1}{M_1^2(E^2 - c_s^2 p^2)}\r]^2 \supset \fr{p^3_{\rm strong}}{E_{\rm strong}} \, k^2 \sim M_1^2 c_s^2 k^2 \qquad\,\,\,  \label{cubic} \\
\parbox[b][12.5mm][t]{20mm}{
\begin{fmfgraph}(35,65) 
\fmfleft{i1} \fmfright{o1}
\fmf{plain}{i1,v,o1} 
\fmf{plain,tension=.4}{v,v}
\fmfdot{v}
\end{fmfgraph}} \!\!\!\!\! &\sim& \int dE d^3 p (M_1^2 p^2) \fr{1}{M_1^2(E^2 - c_s^2 p^2)} \supset \fr{p^3_{\rm strong}}{E_{\rm strong}} \, k^2 \sim M_1^2 c_s^2 k^2\, , \quad  \label{quartic}
\ea
%\end{subequations}
%
where solid lines can stand for both longitudinal and transverse modes, and we pushed the UV cutoffs for loop integrals all the way to the energy and momentum strong-coupling scales.
These results show that the tree-level propagation speeds receive at most an order-one {\it relative} correction,
\be
\Delta c_s^2 \sim c_s^2 \; , 
\label{deltac}
\ee
which means in fact that framids with a small $c_s$ are technically natural. 

Interestingly,  this result is not  a peculiarity of the system at hand, but simply follows from dimensional analysis and perturbativity: small sound speeds are {\it always} technically natural.  
The more general argument goes as follows. 
 In derivatively coupled theories, the dimensionless loop expansion parameter controlling the effects of  quantum fluctuations at a momentum scale  $k$ is   $k/\Lambda$, where $\Lambda$ is the maximal momentum scale at which we can make sense of our effective field theory. (Above, we were calling this $p_{\rm strong}$; for notational simplicity we will now switch to $\Lambda$). Working in units where velocity is dimensionful, by dimensional analysis  we expect quantum corrections from scale $k$ to give
\be
\Delta c_s^2= c_s^2 P(k/\Lambda)
\label{natural}
\ee
where $P$ is a series with coefficients that---if we have properly identified $\Lambda$---are at most $O(1)$. Perturbativity, i.e. $k<\Lambda$, then implies  $\Delta c_s^2 \lsim c_s^2$, as in eq.~(\ref{deltac}). Notice indeed that in deriving eq.~(\ref{deltac}) it was essential to use the explicit value of the cut-off $\Lambda$, which  vanishes like $c_s^{3/2}$ when $c_s\to 0$ with all other Lagrangian parameters held fixed. One may object to the schematic result in eq.~(\ref{natural}),
by noticing there may also appear positive powers of $c'/c_s$, where $c'\gg c_s$ is another velocity. However one is easily convinced that that is not possible, provided  the strong coupling scale $\Lambda$ has been properly identified. The reason is that, with all other terms in the Lagrangian kept fixed, $1/c_s$ controls the strength of the interaction (the strong-coupling scale $\Lambda$ decreases if $c_s$ does): the presence of positive powers of $c'/c_s$ in  eq.~(\ref{natural}), would allow to increase the value of $c_s$ substantially, and thus make the interaction substantially weaker, while remaining in the perturbative regime $k/\Lambda<1$. But that is a contradiction.

The above argument can be made very concrete by considering for instance the most general Lagrangian for a non-relativistic scalar endowed with a shift symmetry\footnote{Our framid's Goldstones, like pions, do not enjoy a shift symmetry. For simplicity we restrict here to shift-symmetric Lagrangians, but  extending our arguments to more general cases is straightforward.} 
\be
S = \int d^3x dt \bigg\{ \frac12 \big( \dot \phi^2 - c^2_s (\vec \nabla \phi)^2 \big) + M^4 F\Big(\frac{\dot \phi}{M^2}, \frac{\nabla \phi}{M^2}, \frac{\partial_t}{M} , \frac{\nabla}{M} \Big)\bigg\}
\ee
where $F$ is a generalized polynomial, not necessarily with $O(1)$ coefficients: the scale $M$ is just a dimensionful unit.

Now, we can get rid of $c_s$ from the quadratic action by defining a rescaled time coordinate $t' = c_s t$. If at the same time we re-normalize the field canonically, ${\sqrt {c_s}} \phi = \phi_c$, we simply obtain 
\be \la{rescaledS}
S = \int d^3x dt' \bigg\{ \frac12 \big( \phi'_c {}^2 - (\vec \nabla \phi_c)^2 \big)  + \frac {M^4 }{c_s}F\Big(\sqrt{c_s} \frac{ \phi'}{M^2}, \frac{1}{\sqrt{c_s}}\frac{\nabla \phi}{ M^2}, c_s \frac{\partial_{t'}}{M} , \frac{\nabla}{M} \Big)\bigg\} \; ,
\ee
where $(\;) ' \equiv \partial_{t'}$.
The numerical coefficients within $F$ and the dependence on $c_s$ will now determine for each interaction term a corresponding  energy/momentum strong coupling scale, $\omega'\sim k\sim \Lambda$, which we do not need to specify. 
Regardless of the details of these interactions, within the perturbative regime, by definition, the kinetic term will receive
at most $O(1)$ corrections. When translating this result back to the original coordinates $(x,t)$, we get eq.~(\ref{natural}).
Eq.~(\ref{natural}) just relies on dimensional analysis, that is the selection rules for independent dilations of time and space. The usual result on  naturalness, or lack thereof, for a relativistic scalar's mass can be stated in the same language. Assuming
the non-linear symmetry  protecting the mass term is broken by some coupling whose loop counting parameter  at the scale $\Lambda$ is $\alpha$, then we expect 
\be
\Delta m^2 = \Lambda^2(c_1\alpha +c_2\alpha^2+\dots)
\ee
compatibly with dimensional analysis and shift symmetry selection rules.

%As a cross check, notice that the same cannot be said for corrections to the mass parameter, if this is not protected by a non-linearly realized symmetry like for our Goldstones: even if one chooses coordinates such that a nonzero tree-level mass is set to one, this can receive huge corrections $\Delta m^2 \propto \Lambda_{\rm UV}^2 $ coming from renormalizable couplings in loops. In our case renormalizable couplings cannot destabilize the propagation speed, because they respect the quadratic action's Lorentz invariance in primed coordinates.
%

The general argument below eq.~(\ref{natural}) implies that a small speed is natural even
when there is a large hierarchy between two propagation speeds in the same system, $c_- \ll c_+$
\footnote{This had better be a technically natural situation, for the excitations of condensed matter systems in the real world are always coupled  to the electromagnetic field at some order in perturbation theory (and to the gravitational one, at some even higher order).}. One quick way to see that is 
 to rescale time using the {\it smaller} speed, i.e. $t' = c_- t$. Notice that, in these units, the propagation speed of the fast modes formally becomes $c_+' = {c_+}/{c_-} \gg 1$.   
%The strong coupling scale is now $\Lambda = M_1 c_<^{5/2}$, and 
The previous argument still applies to loop diagrams that involve only the propagator of the {\it slow} field. In other words, diagrams that only involve slow fields will still give $\mathcal{O}(1)$ contributions to the rescaled action, which means that both propagation speeds will receive $O(1)$ relative corrections, which don't destabilize the hierarchy. For instance $c_+$ will receive an $O(1)$ correction via the renormalization of the coefficient of $\dot \phi_+^2$. Loop diagrams that also involve the {\it fast} field will instead give corrections that are even smaller, being suppressed by at least a factor of $({c_-}/{c_+})^2$ for each fast propagator. In fact, after rescaling time and canonically normalizing the fields, the propagator of the fast field evaluated at the strong coupling scale is 
\be
\l.\fr{1}{E' {}^ 2 - (c_+ / c_-)^2 \, p^2} \r|_{E' \sim p \sim \Lambda} \sim \l(\fr{c_-}{c_+}\r)^2 \fr{1}{\Lambda {}^2}.
\ee
This should be compared with the propagator of the slow field, which instead scales like $1/\Lambda {}^2$ without any additional suppressions. 

The argument above can be easily extended to generic systems with an arbitrary number of fields with different propagation speeds. After rescaling time using the smallest propagation speed and canonically normalizing all fields, one can determine the strong coupling scale $\Lambda$ by looking at the smallest scale that suppresses irrelevant operators. The largest loop corrections to the rescaled action will then come from diagrams that only involve interactions suppressed by $\Lambda$ and propagators of the slowest modes, but such corrections will be at most of $\mathcal{O}(1)$:
%Once again, this means that when we rescale back to the original time coordinate, all propagation speeds (squared) will change by an amount of order of the {\it smallest} propagation speed (squared), and
 {\it hierarchies in propagation speeds are always technically natural}. 
%We should stress  that this result was derived without relying on Lorentz invariance, and therefore it also applies to non-relativistic systems~\cite{Griffin:2013dfa}. 
%

\subsection{\dots but large stresses} \label{stress-che-stress}

Even though it is perfectly natural for framids to have a small speed of sound, there is another sense in which they are, after all, intrinsically relativistic systems, and it has to do with the form of their stress-energy tensor. For type-I framids, we can calculate the stress-energy tensor as usual by varying the action (\ref{lagrangianA}) w.r.t.~an external gravitational field. The calculations are somewhat involved, because one needs to take into account the  $A^\mu A^\nu g_{\mu\nu} = -1$ constraint. The reader can find the details in Appendix~\ref{app_B}, and the final result in eq.~\eqref{timunu_curved}.
%
%Here we report only the final result, which is
%%
%\begin{align} 
%T^{\mu \nu} & = ({\cal L} + \Lambda) \eta^{\mu \nu}\label{timunu}\\
%& - M_2^2 \left[ \d_\rho A^{(\mu} \d^{\nu)} A^\rho -  \d_\rho A^\rho \d^{(\mu} A^{\nu )} - \d^{(\mu} A^{\rho } \, \d^{\nu)} A_{\rho} + A^{(\mu} \d_\rho \d^{\mu )} A^\rho - A^\rho \d_\rho \d^{(\mu} A^{\nu )} \right] \nonumber \\
%&- M_3^2 \left[-g^{\mu \nu} (\d_\rho A^\rho)^2 - g^{\mu \nu} A^\rho \d_\rho \d_\sigma A^\sigma + A^{(\mu} \d^{\nu )} \d_\rho A^\rho\right] \nonumber \\
%&- (M_2^2 - M_1^2) \left[\dot A^\rho \d_\rho A^{(\mu} A^{\nu)} - \dot A^\rho A^{(\mu}  \d^{\nu )} A_{\rho} - \dot A^\mu \dot A^\nu - \d_\rho A^\rho A^{(\mu} \dot A^{\nu)}  \nonumber \right. \\ 
%&\ \ \ \ \ \ \left. + A^\mu A^\nu \d_\rho A^\sigma \d_\sigma A^\rho + A^\mu A^\nu A^\rho \d_\sigma \d_\rho A^\sigma - A^\rho A^\sigma \d_\rho \d _\sigma A^{(\mu } A^{\nu )}\right],  \nonumber 
%\end{align}
%%
%where for notational simplicity we have defined $\dot{A}_\mu \equiv A^\nu \d_\nu A_\mu$, and for completeness we have also included the contribution from a cosmological constant $\Lambda$.  
Note that one could derive the same result  from the Goldstone Lagrangian \eqref{actionI} (or \eqref{effaction} in the case of type-II framids), by modifying the coset construction to include couplings with gravity, as described for instance in~\cite{Delacretaz:2014oxa}. Such a procedure is unambiguous and  perfectly equivalent to the one we have adopted here. 

Even without looking at the explicit form of $T^{\mu\nu}$, it is clear that since each Lagrangian term in \eqref{lagrangianA} involves two derivatives, varying w.r.t~an external gravitational field yields  a stress-energy tensor where each term involves two derivatives, as is indeed the case for the expression \eqref{timunu_curved}. Then, the only non-vanishing contribution to the stress-energy tensor {\it at equilibrium}  is that coming from the cosmological constant, 
\be
\langle T^{\mu\nu} \rangle = \Lambda \, \eta^{\mu\nu} \; ,
\ee 
simply because all other terms contain two derivatives acting on some $A^\mu$ and therefore vanish on the ground state $\langle A^\mu \rangle= \delta^\mu_0$. Hence, we see that type-I framids at equilibrium feature a highly relativistic pressure,
\be
p = - \rho = \Lambda \; .
\ee
The same argument and conclusion apply to type-II framids as well.
The relativistic nature of $T^{\mu\nu}$ makes it hard to imagine how framids could be assembled by handling atoms in a laboratory setting. However, it leaves open the possibility that framids could arise in intrinsically more relativistic situations, like, for instance,  unconventional  phases of QCD. 

Notice that it is actually quite remarkable that the background value of $T^{\mu\nu}$ for framids is Lorentz invariant despite  Lorentz symmetry's being spontaneously broken. The same ``accident" happens for the ghost condensate \cite{ArkaniHamed:2003uy}, but only for a special value of the condensate. Here it is unavoidable, and we believe it deserves further study: apparently there is no selection rule forbidding Lorentz-violating entries in $\langle T^{\mu\nu} \rangle$, yet these vanish. Could a technically similar mechanism be at work in keeping the cosmological constant small in the real world?

%%%%%%%%%%%%%%%%%%%%%%%%%%
%%%%%%%%%%%%%%%%%%%%%%%%%%
\subsection{Thermodynamics, or lack thereof}

The fact that the pressure and energy density of framids are those of a cosmological constant is deeply related to the peculiar 
%has important implications also for the 
thermodynamics of these systems. To begin with, having thermodynamic degrees of freedom at all seems a fundamental property of  ordinary condensed matter systems: usually we are able to slightly change their state by exerting  pressure or strain  on them, by changing their temperature or chemical potential, etc. Thermodynamic degrees of freedom come in conjugated pairs, like for instance pressure and volume $(p,V)$, temperature and entropy $(T,S)$, chemical potential and charge $(\mu, Q)$, etc. The intensive variable in the pair ($p, T, \mu, \dots$) can be viewed as a thermodynamic control parameter. The extensive variables $V, S, Q, \dots$ can also be traded for the corresponding densities: $1$ for $V$, $s=S/V$, $n=Q/V$, etc.
%In fact, in most cases thermodynamics is part of the microscopic definition of the system, for instance via the partition function.

In our SSB/EFT language, we can keep track of thermodynamic  variables for ordinary condensed matter systems in two equivalent ways:
\begin{enumerate} 
\item 
{\it At the level of the symmetry breaking pattern.} When the unbroken translations $\bar P_\mu$ are non-trivial linear combinations of the Poincar\`e generators and of internal symmetries,  thermodynamic control parameters {\it do} appear in their definitions. For instance, for superfluids, the unbroken time translation operator $\bar P^0 = P^0 - \mu Q$ involves  an arbitrary dimensionful parameter $\mu$ that can be interpreted as the chemical potential.\footnote{In the classification of sect.~\ref{SSB} all such parameters have been omitted to simplify the notation. For ordinary fluids we have $\bar P^i = P^i - s^{1/3} Q^i $, where $s$ is the entropy density \cite{Dubovsky:2011sj}, while for solids we have to allow for equilibrium shear deformations as well,  $\bar P^i = P^i - A^i {}_a Q^a$, where $A $ is a matrix related to the strain tensor \cite{Landau_solids}.} 
\item 
{\it At the level of the effective theory.} Thermodynamic control parameters describe the  ``moduli space" of solutions
satisfying space-time homogeneity. 
For instance (see case 2, sect. 2.1) a superfluid field theory has the family of solutions $\psi(x)  = \mu t$, parametrized by the chemical potential $\mu$. A change in chemical potential corresponds to exciting a suitable configuration $ \pi(x)  = \delta \mu \cdot t$ of the Goldstone boson $\pi$ describing  small fluctuations of $\psi$ around its vev, $\psi(x) = \mu t + \pi(x)$~\cite{Nicolis:2011pv}.  Analogous considerations can be carried out for solids and liquids as well.
\end{enumerate}
Notice that property 2 above is not the usual statement that constant Goldstone field configurations can make one move from a ground state to an equivalent one, related to the first by a symmetry transformation. Rather, here there is a continuum of {\it physically inequivalent} solutions---for instance, they have different energy density and pressure---and the Goldstone configurations that interpolate between them have nontrivial space-time dependence. 
Properties 1 and 2 are equivalent, in that one can prove in broad generality that the moment one has an unbroken combination of a translation operator and of an internal charge, for instance $\bar P^0 = P^0 - \mu Q$, one can explicitly construct a Goldstone coherent state that shifts the value of $\mu$, thus making  $\bar P^0 = P^0 - (\mu + \delta \mu) \, Q$ the new unbroken combination \cite{Nicolis:2011pv}. This property  also connects to the scaling of the scattering amplitudes with energy  for such systems: $\pi=\delta \mu t$ is a solution of the equations of motion provided that the Langrangian only depends on $\partial_\mu\pi$, in which case amplitudes scale like $E^4$.

It is intuitive that thermodynamic control parameters should be associated with a ``moduli space" of inequivalent 
homogeneous and isotropic solutions solutions. The inequivalence of the solutions corresponds to the inequivalence of the boundary conditions, and the latter can be associated with the action of an external control parameter. The inequivalence
of the solutions, in particular their coordinate dependence, also implies a redefinition of the unbroken translation generators, hence the dependence of the latter on the control parameters themselves. However we have no general theory for the above. It is just a fact that holds for ordinary condensed matter systems---all the systems not involving extra space-time symetries in sect.~\ref{SSB}, apart from framids.

%We should emphasize that it is not clear to us at present why thermodynamical transformations should show up in these ways in our formalism. In particular, why general thermodynamical variables should have anything to do with SSB, at the level of the algebra or of the Goldstone EFT. It is just a fact that holds for ordinary condensed matter systems---all the systems not involving extra space-time symetries in sect.~\ref{SSB}, apart from framids.

It is also evident that the presence of thermodynamic control parameters is directly associated with the non-triviality of the energy momentum tensor ($T_{\mu\nu}\not \propto \eta_{\mu\nu}$ ) over homogenous configurations. For instance, by considering a general superfluid field theory one can easily show that over homogeneous configurations one has
\be
p+\rho= n\mu
\ee
where $n=J_0$ is the charge density. The above equation is nothing but the usual thermodynamic relation (with $s$ and $T$ entropy density and temperature respectively)
\be
p+\rho= sT+n\mu
\label{termorelativity}
\ee
evaluated at zero temperature. Moreover eq.~(\ref{termorelativity}) does hold in the field theory describing a relativistic fluid~\cite{Dubovsky:2011sj}. While for general solids one can easily prove
\be
p_{ij}+\rho \delta_{ij}\equiv T_{ij}+\rho \delta_{ij}=-J_i^a A_j^a
\ee
where $J_a^i$ and $A^i {}_a$ are respectively the current density and the conjugated control parameter (associated with the strain).

Let us now focus on framids. Neither property 1 or 2  holds for them.
As for property 1, it is interesting to note that framids are the only condensed matter systems in our classification that  do not contain continuous adjustable parameters in their symmetry breaking pattern. Framids of type I do not possess additional internal symmetries at all, and thus their unbroken translation and rotation generators coincide with the original,  Poincar\'e ones. Framids of type II possess an internal $SO(3)$ symmetry that mixes with spatial rotations to generate an unbroken diagonal combination $\bar J_i = J_i + \tilde Q_i$.
%, where $J_i$ are the generators of rotations and $\tilde Q_i$ those of the internal $SO(3)$. 
However, due to the non-Abelian nature of these groups, it is easy to convince oneself that such an unbroken combination does not allow any adjustable parameter in it.

As for property 2, one can check that framids do not possess a moduli space of inequivalent homogeneous and isotropic solutions.
That is simply because the Lagrangian  does not depend on just the Goldstone derivatives, but also on the Goldstones themselves. This is associated with the absence of  abelian generators mixing with translations, that is to say, with the fact that   Poincar\'e translations are unbroken.

To be more explicit, one can check if considering non-trivial configurations $A^\mu(x)$, could give rise to a homogeneous, stationary energy momentum tensor of the form
\begin{equation} \label{diagonal}
T^{\mu \nu} = {\rm diag}(\rho, p, p, p)\, ,
\end{equation}
such as what we expect for an isotropic medium at equilibrium. By careful inspection of the formula for $T^{\mu \nu}$ given in the appendix,~\eqref{timunu_curved}, one can show that this is in fact not possible. The part of $T^{\mu \nu}$ proportional to $c_2$ can be put in the form~\eqref{diagonal} by choosing $\partial_\mu A^\mu = {\rm const.}$ ({\it e.g.},  $A^i = x^i, A^0 = \sqrt{1+ x^i x^i}$). In particular, this part gives again a contribution analogous to a cosmological constant, with $p = -\rho$.  However, the same choice for $A^\mu$ makes the other terms inhomogeneous.
%and isotropic.  

We thus conclude  that framids do not seem to possess thermodynamical properties in any standard sense, or at least none that is visible at the level of the symmetry breaking pattern or of the low-energy EFT for the corresponding Goldstones: they seem to possess only {\it one} equilibrium state, and not the continuum associated with more ordinary thermodynamical systems\footnote{The framid equilibrium state can be boosted of course, thus formally yielding a continuum of equilibrium states, but these all have the same physical properties, being related to one another by  symmetry transformations.}.

%%%%%%%%%%%%%%%%%%%%%%%%%%%%%%%%%%%%%%%%%%%%
%%%%%%%%%%%%%%%%%%%%%%%%%%%%%%%%%%%%%%%%%%%%

\section{A first look at extra-ordinary stuff}\label{sec:galileids}

In Sect.~\ref{SSB} we left open the possibility that, for certain condensed matter systems, the residual homogeneity and isotropy featured at low energies could be due to unbroken combinations of Poincar\'e generators and other  {\it spacetime} symmetries, that is, symmetries that do not commute with the Poincar\'e group itself. These additional spacetime symmetries are the defining feature of what we will call {\it extra-ordinary} (as opposed to ordinary) condensed matter systems. Extending our classification to all such systems is too daunting a task to be addressed here in full generality\footnote{For instance, one could embed 4D Poincar\'e into the isometry groups of higher dimensional spaces, and start playing with branes shaped in such a way as to preserve the desired unbroken symmetries.}. Instead, we will limit ourselves to the study of those spacetime symmetries that are needed to complete the classification of sect.~\ref{SSB}. We refer, in particular, to  those symmetry breaking patterns, cases 3 and 5, that cannot be realized by supplementing the Poincar\'e group with internal symmetries only. For future reference, let's remind the reader that case 3 corresponds to having unbroken translations and rotations
of the form
\be \la{case3}
\bar P^0 = P^ 0\; , \qquad \bar P^ i = P^i + \beta \, Q^i \;, \qquad \bar J = J^i \; , \qquad \mbox{(case 3),}
\ee
where the $Q^i$'s are the extra generators we are after. Likewise, for case 5:
\be \la{case5}
\bar P^0 = P^ 0 + \alpha \, Q \; , \qquad \bar P^ i = P^i + \beta \, Q^i \;, \qquad \bar J = J^i  \; , \qquad \mbox{(case 5).}
\ee
Notice that, compared to our analysis in Sect. \ref{SSB}, we have now explicitly introduced the control parameters $\alpha$ and $\beta$. In light of our discussion on the thermodynamics of framids, this suggests that the above scenarios will be endowed with non-trivial thermodynamic properties. Nevertheless, we will discover other reasons why these two cases are pathological and cannot describe condensed matter systems that are physically realized in Nature. In particular, we will find that, at least in their simplest realizations, these scenarios   are either plagued by instabilities, or have non-homogeneous stress-energy tensors. We should stress however that extra-ordinary systems can appear in any of the eight scenarios discussed in section \ref{SSB}, and we know for certain that in some of these cases they correspond to physically sensible systems. We will elaborate further on this point in the final section of this paper, but we leave for future work a more thorough analysis of extra-ordinary systems.

% these pathologies  seem to be just a consequence of the particular symmetry breaking patterns (\ref{case3}) and (\ref{case5}), and are not a general feature of all extra-ordinary systems. 

\subsection{Minimal symmetry realization} \la{subs5.1}

The unbroken generators $\bar P^ i$ in eqs. (\ref{case3}) and (\ref{case5}) have the correct transformation properties under rotations only if the $Q^i$'s transform like 3-vectors. Then, for consistency these generators must belong to a multiplet that transforms according to  some representation of the Lorentz group. The most economical possibility is to assume that the generators $Q$ and $Q_i$ make up a Lorentz 4-vector $Q_\mu$ of generators that satisfy the following algebra:
\be \la{comm}
[  Q_\mu, Q_\nu ] = [  Q_\mu, P_\nu ] = 0 \; , \qquad \qquad \l[J_{\alpha\beta},Q_\gamma \r] = i ( \eta_{\gamma\alpha} Q_\beta  - \eta_{\gamma\beta} Q_\alpha) \; .
\ee
Then, the scenario 5 in which all original translations are broken can be easily realized by a vector field $B_\mu$ that shifts under the action of the $Q_\mu$'s, i.e. $ B _\mu \to B_\mu + c_\mu$ with $c_\mu$ constant, and acquires an expectation value
\be \la{Bvev}
\langle B_\mu \rangle =  \alpha t \, \delta_\mu^0 + \beta x_i \, \delta_\mu^i \; .
\ee
Notice that because of the shift symmetry, the Lagrangian will only depend on derivatives of $B_\mu$, and therefore (\ref{Bvev}) will be a solution for all real values of $\alpha$ and $\beta$, as befits their interpretation as thermodynamic parameters.  The scenario 3 in which both $P_0$ and $Q_0$ remain unbroken is considerably more complicated to implement at the level of fields. It can be realized for instance via a reducible representation of Lorentz, made up of a vector $C_\mu$ and a scalar $\vphi$ that transform under $Q_\mu$ as 
\begin{subequations} \la{transfQ}
\ba
\vphi &\to& \vphi + 2 b^\mu C_\mu \\
C_\mu  &\to& C_\mu + b^\nu(\d_\mu C_\nu +\d_\nu C_\mu) - \tfrac{1}{2} b^\nu \d_\nu \d_\mu \vphi \; ,
\ea
\end{subequations}
and that acquire the expectation values
\be \label{penchide}
\langle \vphi \rangle = \beta \vec x^{\, 2}, \qquad \quad \langle C_\mu \rangle =  \beta x_i \, \delta_\mu^i \; .
\ee

Writing down the most general low-energy effective action for these order parameters can be rather cumbersome, especially in the second case where it is not obvious how to systematically classify invariants under the symmetry transformations (\ref{transfQ}).\footnote{Notice that the transformations (\ref{transfQ}) act like a translation on the combination $\d_\mu C_\nu +\d_\nu C_\mu - \d_\mu \d_\nu \vphi$, which readers familiar with the St\"uckelberg formulation of massive gravity will easily recognize \cite{ArkaniHamed:2002sp}. As such, any Lagrangian that is built out of this combination and is invariant under translations will also be invariant under the transformations generated by the $Q_\mu$'s. However, we don't have a proof that this combination is the only one that is allowed to lowest order in the derivative expansion.} For this reason, we find it more convenient to study these scenarios using the coset construction reviewed in sect.~\ref{sec:coset}. 
%Although it could be also interesting to carry out the same analysis starting from the order parameters, we are confident that the results would be the same. 
%
%
Let us start by considering scenario 5, which from the coset viewpoint is more general. As we will see, scenario 3 can be recovered as a special case. The starting point is the coset parametrization, which we choose to be 
\begin{equation}
\Omega = e^{i x^\mu \bar P_\mu} e^{i \xi^\mu(x) Q_\mu} e^{i \eta^i(x) K_i},
\end{equation}
where  $\xi^\mu(x)$ and $\eta^i(x)$ are the Goldstone fields. 
% for the broken charges $Q_\mu$, while the $\eta^i(x)$ are the Goldstones of the broken boosts. 
Using the Poincar\'e algebra together with the commutation relations (\ref{comm}), we can calculate the Maurer-Cartan form $\Omega^{-1} \partial_\mu \Omega$ and cast it in the form
\be \la{MCform35}
\Omega^{-1} \d_\mu \Omega \, = i e_\mu{}^\nu \l( \bar P_\nu + D_\nu \xi^\rho Q_\rho + D_\nu \eta^i K_i + {\cal A}_\nu^i \bar J_i \r).
\ee
Notice however that  these systems can exhibit the inverse Higgs mechanism, which would make the boost Goldstones $\eta^i$ redundant. This can be easily deduced by using the same criterion that we have  used throughout the paper, i.e.~by noticing that the commutators between unbroken translations and boosts contain the broken generators $Q_\mu$,
\begin{align}
[ K_i, \bar P_0] &= i [\bar P_i + (\alpha-\beta) Q_i ] \\
[K_i, \bar P_j] &=  i \delta_{ij} [\bar P_0 + (\beta - \alpha) Q_0]\, .
\end{align}
Within the coset construction, these redundant Goldstone modes can be eliminated by imposing the inverse-Higgs constraints, i.e.~by setting to zero certain covariant derivatives and solving for the redundant Goldstones in terms of all the other ones. In our case, the most general inverse Higgs constraint we can impose to lowest order in derivatives is~\cite{Ivanov:1975zq}
\be
c_1 D_i \xi^0 + c_2 D_0 \xi^i = 0.
\ee
where $c_1$ and $c_2$ are arbitrary coefficients, not necessarily constant, but invariant under the symmetries (cf.~\cite{Endlich:2013vfa}). Notice that these conditions preserve all the symmetries because $D_i \xi^0$ and $D_0 \xi^i$ transform covariantly. 

Different choices for the values of $c_1$ and $c_2$ yield physically equivalent effective Lagrangians for the remaining Goldstones.
This can be understood by recalling that for certain symmetry breaking patterns, the inverse-Higgs constraints can be interpreted as gauge-fixing conditions for certain gauge-redundancies associated with the Goldstone parameterization of the order parameter's fluctuations \cite{Nicolis:2013sga, Endlich:2013vfa}. This is clearly the case for the $B_\mu$ implementation of our symmetry pattern, 
eq.~\eqref{Bvev}: $B_\mu$ only has four independent components, so their parametrization in terms of seven Goldstone fields must be redundant. Then, different values of $c_1$ and $c_2$ correspond to different gauge choices for the same physical system---all of which remove the redundant Goldstones in a way that is consistent with the global symmetries. 

Without loss of generality, we can then set  $D_0 \xi^i = 0$, i.e.~$c_1 = 0$ 
\footnote{As a check, we performed the analysis below for generic values of $c_1$ and $c_2$, obtaining the same results as below. For simplicity, we will not report that general analysis here.}.
Using the fact that the covariant derivatives of the $\xi^\mu$'s read
\begin{align} \label{covariant}
D_\mu \xi^\nu &  = (\alpha - \beta) [\delta_\mu^i \delta_i ^\nu - (\Lambda^{-1})_\mu {}^{i} \Lambda_i {}^{\mu} ] + (\Lambda^{-1})_\mu {}^{\beta} \Lambda_\gamma {}^{\nu}  \partial_\beta \xi^\gamma \, ,
\end{align}
where  $\Lambda_\mu {}^{\nu}$ is once again a boost with rapidity $\vec{\eta}(x)$ and  $\Lambda_0 {}^{i} = \Lambda_i {}^{0} = \eta^i + \mathcal{O} (\vec{\eta}^{\, 3})$, we obtain the following relation to linear order in the fields:
\begin{equation} \la{IHsol}
\eta^i \simeq \frac{\dot \xi^i}{\beta - \alpha} \, .
\end{equation}
The covariant derivatives $D_\nu \eta^i \sim \d_\nu \dot \xi^i$ are then of higher order in the derivative expansion and become negligible at low energies. Following the procedure outlined in section \ref{sec:coset}, the most general low-energy Lagrangian that is invariant under all the symmetries can be obtained by including all possible contractions of the covariant derivatives $D_0 \xi^0, D_i \xi^0$ and $D_i \xi^j$ that are manifestly invariant under rotations. The only terms that contribute to the quadratic Lagrangian for the Goldstones are 
\begin{equation} \la{Lvecgal}
{\cal L} \simeq \lambda_1 D_0 \xi^0 + \lambda_2 (D_0 \xi^0)^2 + \lambda_3 D_i \xi^i + \lambda_4 D_i \xi_j D^i \xi^j + \lambda_5 D_i \xi_j D^j \xi^i + \lambda_6 (D_i \xi^0)^2 \, ,
\end{equation}
where, after plugging the result (\ref{IHsol}) in the covariant derivatives~\eqref{covariant} and performing the rescaling $\xi^\mu \rightarrow (\beta - \alpha) \xi^\mu$, we have 
\begin{subequations} \la{problem}
\begin{align}
D_0 \xi^0 & = \dot \xi^0 - \dot \xi^j \partial_j \xi^0\, + \mathcal{O}(\xi^3),\\
D_j \xi^0 & = \partial_j \xi^0 + \dot \xi^j + \dot \xi^i \d_i \xi_j + \partial_j \xi^i \dot \xi_i + \mathcal{O}(\xi^3) , \\
D_j \xi^i & = \partial_j\xi^i + \dot \xi^i \partial_j \xi^0 + \mathcal{O}(\xi^3) \, \label{maporca}.
\end{align}
\end{subequations}

It is now easy to see that the kinetic term for $\xi^i$ and the gradient term for $\xi^0$ both come from the last term in the Lagrangian (\ref{Lvecgal}). Therefore they always have the same sign, which means that if require that the $\xi^i$ fields are not ghost-like, we inevitably end up with gradient instabilities for the $\xi^0$ field, and viceversa. We conclude that this minimal realization of  scenario 5 is inconsistent.\footnote{In principle one may wonder whether the mixing terms could affect this conclusion. However, a straightforward Hamiltonian analysis is sufficient to establish once and for all that such terms are not able to restore positivity.}

A similar conclusion applies to  scenario 3 as well. Formally, we can derive the corresponding effective Lagrangian for this scenario by setting $\alpha=\xi^0 =0$ and neglecting all the covariant derivatives of the  $\xi^0$ Goldstone, because in this case the corresponding charge $Q_0$ is unbroken. The covariant derivative $D_j \xi^i $ is the only ``building block'' available to lowest order in derivatives, because the $D_0 \xi^i$ have been set to zero by the inverse Higgs constraints. A quick glance at eq.~\eqref{maporca}  is then sufficient to reveal the problem: it is impossible to write a kinetic term for the $\xi^i$ fields at lowest order in derivatives, because $D_j \xi^i$ does not contain a quadratic piece of the form $\dot \xi_j \dot \xi^i$. The dynamics is then controlled by higher time-derivative terms such as $(D_0 \eta^i)^2 \sim  (\ddot \xi^i)$ which inevitably lead to ghost instabilities. Thus, this minimal approach to  scenario 3 is also inconsistent. 

It is worth pointing out that, if it weren't for the instabilities, the two models above could correspond to fairly standard condensed matter systems:  they have a stress-energy tensor that (1) is homogeneous on the background, (2) has $\rho + p \neq 0$, and (3) depends on some thermodynamical control parameters ($\alpha$ and $\beta$) that can be varied continuously by exciting suitable configurations of the Goldstone fields. These properties can be deduced more easily using the order parameters. For instance, the low-energy effective action for the order parameter $B_\mu$ of case 5 contains all Lorentz invariant combinations of first derivatives $\d_\mu B_\nu$, whose expectation value on the background (\ref{Bvev}) is constant and breaks Lorentz, i.e.~$\langle \d_\mu B_\nu \rangle = \alpha \delta_\mu^0 \delta_\nu^0 + \beta \delta_\mu^i \delta^i_{\nu}$. This means that the stress-energy tensor is homogeneous on the background, and that in general $\rho$ and $p$ are different. Finally, it is easy to see that a Goldstone configuration such as $\xi_\mu = \delta \alpha \, t \delta_\mu^0 + \delta \beta \, x_i \delta_\mu^i$ effectively corresponds to a change in the parameters $\alpha$ and $\beta$, which therefore can be varied continuously. Similar considerations apply to the stress-energy tensor of the realization \eqref{penchide} of case 3.

\subsection{Minimal particle-content realization: the galileids}

We will now show that the instabilities encountered above can be circumvented by adding a single generator to the algebra, but the price to pay is that the expectation value of the stress energy tensor is no longer homogeneous on the background. 
% {\bf RP: how about the thermodynamic variables?} 
More precisely, we will now modify the algebra (\ref{comm}) by adding a generator $D$ that satisfies the following commutation relations:
\be \label{gal-al1}
[D,Q_\mu]  = [D, P_\mu]  = [D, J_{\mu\nu}]  = \ 0\, , \qquad \quad  [Q_\mu, P_\nu] \  = \  2 i \eta_{\mu \nu} \, D\, .
\ee

The algebra above  defines the symmetries of galileon theories \cite{Nicolis:2008in}.  These involve a scalar field $\phi$
enjoying a generalized shift symmetry of the form
\be \label{galileon}
\phi(x) \to \phi(x) + c + b_\mu x^\mu \; ,
\ee
where $c$ and $b_\mu$ are constant. To make contact with the generators above, the shift by $c$ is generated by $D$, the shift by $b_\mu x^\mu$ is generated by $Q^\mu$. One can check straightforwardly that the algebra is precisely the one in eq. (\ref{gal-al1}). %Incidentally, the reason why the symmetries generated by $Q_\mu$ and $D$ can be implemented non-linearly on a {\it single} scalar field is again because of the inverse Higgs mechanism: since $[Q_\mu, \bar P_\nu] \  = \  2 i \eta_{\mu \nu} \, D$, the Goldstones of $Q_\mu$ are redundant and can be expressed in terms of derivatives of the single Goldstone of $D$.

Let us start by considering how  case 3 can be implemented in the context of the galileon algebra. Interestingly, it is possible to show that the algebra (\ref{gal-al1}) is actually the only possible extension of the algebra (\ref{comm}) that is compatible with case 3---see Appendix \ref{appa} for more details. As to the SSB pattern that we are after, in order to preserve `unprocessed' rotations and time-translations according to (\ref{case3}), we need to consider a background solution for $\phi(x)$ of the form
\be
\langle \phi(x) \rangle  = f(|\vec x|^2) \; .
\ee
Then, in order to preserve a linear combination of $P^i$ and $Q^i$, we need
\be
\langle \phi(x) \rangle = \sfrac 12 \beta |\vec x|^2 \; ,
\ee
with constant $\beta$, in which case the unbroken combination is $\bar P^i = P^i + \beta \, Q^i$: a translation can be compensated by a galilean shift, thus yielding a new form of homogeneity.

Is this a solution of the field equations?
Because of the symmetries, the field equations involve {\it at least} two derivatives on each $\phi$. This means that a quadratic configuration of the form above, once plugged into the field equations will yield an algebraic equation for $\beta$, which is in fact a polynomial of at most  third order \cite{Nicolis:2008in}. For generic (if not for all) choices of the Lagrangian coefficients such an equation will have a real solution, which then identifies a background with the right symmetries. 

We call this system a {\it type-I galileid}. Clearly, since the original Lagrangian involves a single scalar degree of freedom, the system features a single Goldstone excitation $\pi(x)$,
\be
\phi(x) = \sfrac 12 \beta |\vec x|^2 + \pi(x) \; ,
\ee 
which can be thought of as that associated with the shift generator $D$.
This is consistent with the existence of six inverse-Higgs constraints, which can eliminate the Goldstones of $K^i$ and of $Q^i$ in favor of $\pi$ and its derivatives, as allowed by the commutation relations
\be
[\bar P^0, K^i] = \bar P^i - \beta Q^i \; , \qquad  [Q^j , \bar P^i] = 2i D \,  \delta^{ij} \; .
\ee

The generalization to case 5 is straightforward. The configuration
\be
\phi(x) = \sfrac 12 \big( \beta |\vec x|^2  -  \alpha t^2 \big)
\ee
preserves
\be
\bar P^0 = P^ 0 + \alpha Q^0 \; , \qquad \bar P^ i = P^i + \beta Q^i \;, \qquad \bar J = J^i \; ,
\ee
which have the right algebra for space-time translations and spatial rotations, thus realizing the desired symmetry breaking pattern. Once plugged into the galileon's field equation, the configuration above yields a single polynomial equation for two variables---$\alpha$ and $\beta$. So, at least in some finite range of real values for $\beta$, we expect a continuum of real solutions with $\alpha = \alpha(\beta)$. Like for case 3 above, there is here a single Goldstone excitation,
\be
\phi(x) = \sfrac 12 \big( \beta |\vec x|^2  -  \alpha (\beta) t^2 \big) + \pi(x) \; ,
\ee
in agreement with the existence of seven possible inverse Higgs constraints associated with the commutation relations
\be
[\bar P^0, K^i] = \bar P^i +(\alpha - \beta) Q^i  \; , \qquad [Q^\nu, \bar P^\mu] = 2 i  D \,\eta^{\mu\nu} \; .
\ee 
We call such a system {\it type-II galileid}. 
%
%After all inverse Higgs constraints have been applied, type I and II galileids feature one Goldstone field each: the one associated with the shift generator $D$. However, $D$ is a central charge in the galileon algebra~\eqref{gal-al1}, and can thus be consistently set to zero. This particular choice would generate a different realisation of the symmetry breaking pattern, in which the Goldstone field of $D$ is absent, and those associated with the $B^\mu$ cannot be reabsorbed by any inverse Higgs mechanism. In particular, one would expect in this case to have 3 and 4 Goldstone fields for cases 3 and 5 respectively. However, as we show in sect.~\ref{curse}, but such systems contain incurable pathologies.  
%
Notice that there is an interesting limit of the type-II galileid in which boosts are {\it not} broken. It corresponds to a configuration of the form above with $\beta = \alpha$,
\be \label{LI_galileid}
\phi(x) = \sfrac 12  \alpha \,  x_\mu x^\mu + \pi(x) \; ,
\ee
which is a solution provided the Lagrangian coefficients obey certain inequalities, and describe the sub-horizon geometry of deSitter-like solutions in modified-gravity theories~\cite{Nicolis:2008in}.

The Lagrangian terms for the galileon that are most relevant at low-energies have the schematic form \cite{Nicolis:2008in}
\be \la{galileons}
{\cal L}_n \sim \partial \phi \, \partial \phi \, (\partial \partial \phi)^{n-2} \; ,
\ee
and are invariant under galilean shifts only up to a total derivative. Notice in particular that  ${\cal L}_2$ is a standard kinetic term for the field $\phi$. In the absence of this term, the dynamics would be controlled by the exactly invariant quadratic term $(\partial \partial \phi)^{2}$ which would lead to ghost-like instabilities. Thus, galileids can be ghost-free because an ordinary kinetic term is invariant under all the symmetries, although only up to a total derivative.\footnote{The coset constructions used in the previous sections yield terms that are  invariant under all the symmetries exactly, and not just up to a total derivative. Thus, one may wonder whether terms in the latter class---which are known as {\it Wess-Zumino terms}---could also be used to eliminate ghosts in the realizations studied in sect.~\ref{subs5.1}. We have explored this possibility but concluded that there are no Wess-Zumino terms that can provide a healthy kinetic term for those~systems.}

Once expanded about the backgrounds above, the terms (\ref{galileons}) yield interactions for the Goldstone excitations that are much ``softer" (at low energies) that those of more standard condensed matter systems (solids, etc.), since they involve on average more than one derivative per field. In particular, the $2\to 2$ scattering amplitude scales like $E^6$ rather than $E^4$. We remind the reader that for framids the low-energy scaling is $E^2$. Galileids are thus ``on the other side" of conventional condensed matter compared to framids, both in terms of the complexity of the associated symmetry breaking patterns, and in terms of the low-energy scaling of scattering amplitudes.

We can use the galileon algebra to implement other cases of our classification of sect.~\ref{SSB} as well, replacing  some of the internal generators with Galilean shifts. We go through a number of examples along these lines in Appendix \ref{general galileids}.
We should also mention that it is  possible in principle to generalize the galileon shifts \eqref{galileon} to involve higher powers of $x^\mu$ as well. For instance:
\be
\phi(x) \to \phi(x) + c + b_\mu x^\mu + d_{\mu\nu} x^\mu x^\nu \; .
\ee
These generalizations have been recently studied in some detail in ref.~\cite{Hinterbichler:2014cwa}. The problem with these higher order symmetries is that the equations of motion---in order to be invariant---need to involve more than two derivatives per field, and this generically leads to ghosts (i.e., negative energy states). The only case that has a chance of being physically well-behaved is one where there are never more than two {\it time}-derivatives on any field. Ref.~\cite{Hinterbichler:2014cwa} has analyzed this possibility for Lorentz-breaking systems. It would be interesting to extend the analysis to our framework as well, where Lorentz invariance is broken only spontaneously. Can one have a system in which the higher order time-derivatives always act on the background configuration and never on the Goldstone excitations? We leave this technical question for future work, and move on to ponder whether galileids can be realized in Nature.

%%%%%%%%%%%%%%%%%%%%%%%%%%%%%%%%%%%%%%%%%%%%
%%%%%%%%%%%%%%%%%%%%%%%%%%%%%%%%%%%%%%%%%%%%
\subsection{Do Galileids exist?} \label{do-they}

Galileids are based on the galileon effective theory, which is notoriously a dubious theory. Its most worrisome  theoretical peculiarities---if not necessarily pathologies---are the existence of superluminal excitations about certain backgrounds, and the unusual softness of scattering amplitudes at low energies, ${\cal M}_{2 \to 2} \sim E^6$ \cite{Adams:2006sv}.  Despite not being inconsistencies of the low-energy effective theory itself, the former obstructs the embedding of the effective theory into a UV-complete theory obeying the standard causal structure of local relativistic  QFTs
\footnote{See however \cite{deRham:2013hsa, Creminelli:2013fxa} for recent twists in this story.}, 
while the latter violates standard dispersion rules following from Lorentz-invariance and $S$-matrix analyticity extrapolated to arbitrarily high energies. 

Notice however that both of these objections could be irrelevant for galileids. The reason is that galileids are formally derived from the galileon effective theory expanded about certain Lorentz-violating solutions, but there is no guarantee that the same effective theory that describes the physics of galileids can be extrapolated to very different backgrounds, for instance the Poincar\'e-invariant one with $\phi=0$.  This is completely analogous to what happens for other condensed matter systems, say a superfluid, where the effective theory for the phonon field $\pi$ is conveniently parameterized in terms of the Lorentz scalar $\psi(x) = \mu t + \pi(x)$, but clearly in general cannot be extrapolated to the Poincar\'e-invariant phase with vanishing chemical potential, $\langle \psi \rangle = 0 $, where the superfluid is gone! Then, the use of the galileon EFT for galileids should conservatively be  thought of  as just an analogous technical shortcut, to encode the spacetime symmetries acting on the galileid's Goldstone excitations  in a simple fashion. In this case, it is entirely possible that the galileon backgrounds formally featuring superluminal excitations are far (in field space) from the galileid background, and thus lie outside the regime of validity of the galileid's Goldstone effective theory. Similarly, one cannot apply relativistic dispersion relations directly to scattering processes involving the galileid's Goldstones, which propagate on a Lorentz-violating background. One could apply them to scattering processes about the Poincar\'e invariant background, $\langle \phi \rangle =0$, but that background might not be covered by the galileid's effective theory.

There are however other peculiarities that make galileids stand out compared to other condensed matter systems. Consider the galileon's stress energy tensor. In terms of fields and derivatives, it has the same schematic form as  the Lagrangian terms it comes from, eq.~\eqref{galileon}: 
\be
T_{\mu\nu} \sim \partial \phi \, \partial \phi \, (\partial \partial \phi)^{n-2} \; .
\ee
When evaluated on a galileid background---which is quadratic in coordinates, $\phi \sim x^2$---it reduces to
\be
\langle T_{\mu\nu} (x) \rangle_{\rm galileid} \sim x^2 \; ,
\ee
that is, is {\it not} translationally invariant!
The reason is that the galileid background is invariant only under the combined action of translations and galilean shifts. A generic local operator that---like $T_{\mu\nu}(x)$---is not invariant under galilean shifts, will not  have an expectation value that is invariant under space-time translations
\footnote{This is to be contrasted with what happens with other condensed matter systems that break translations, like solids for instance. There, the background is invariant under a combination of translations and internal shifts. However, the stress-energy tensor operator is manifestly invariant under internal shifts, and so its expectation value on the background is invariant under translations.}. Notice that a Lagrangian with {\it exact} Galilean symmetry would be a function of $\d \d \phi$ and  would consequently have a homogeneous stress-energy tensor on our background. However, such a Lagrangian would be obviously plagued by ghosts. This connection between ghosts and homogeneity of the energy-momentum tensor, which we have already encountered in section \ref{subs5.1}, seems to be a recurrent theme for systems in classes 3 and 5. 

A related problem concerns the coupling of the galileon to gravity. The galileon shift
\be
\phi(x) \to \phi(x) + b_\mu x^\mu
\ee
cannot be straightforwardly extended to curved space-time, because there $x^\mu$ is not a covariant object anymore. One could try to bypass this by imposing a constant shift symmetry on the {\it derivative} of $\phi$,
\be
\partial_\mu \phi \to \partial_\mu \phi + b_\mu \; ,
\ee
but the only meaningful notion of constant $b_\mu$ in curved space-time is that of vanishing covariant derivatives, $\nabla_\mu b_\nu(x) = 0$, and a generic space-time does not admit any such covariantly constant vector fields
\footnote{If such a $b_\mu(x)$ exists, it must be a Killing vector: $\nabla_\mu b_\nu + \nabla_\nu b_\mu = 0$. Hence, a generic space-time with no isometries cannot support any covariantly constant $b_\mu$. On the other hand, one can successfully generalize the galileon symmetry to  maximally symmetric spacetimes \cite{Goon:2011qf, Burrage:2011bt}, but of course maximal symmetry is gone as soon as gravitational perturbations are taken into account.}. So, straightforwardly coupling the galileon to gravity breaks the galileon symmetry. 

Given these difficulties, it is somewhat ironic that the galileon has emerged as part of gravity itself in theories that modify general relativity in the infrared: in the DGP model \cite{Dvali:2000hr} as a 4D brane-bending mode of 5D gravity \cite{Luty:2003vm, Nicolis:2004qq}, and in massive gravity \cite{deRham:2010ik} as the helicity-zero component of a massive graviton \cite{ArkaniHamed:2002sp, deRham:2010kj}. There, the gravitational couplings of the galileon are certainly not that of a scalar, simply because in these theories there is no scalar degree of freedom to begin with: only in the high-energy regime---at distances much shorter than the IR modification scale---is there an approximately scalar degree of freedom, in analogy with the equivalence theorem for massive gauge bosons. The couplings of this degree of freedom to the other components of the gravitational field vanish in this high-energy regime (the so-called decoupling limit), and there is no regime in which one has a gravitationally coupled scalar. 

Given all of the above, it is thus conceivable that galileids could show up as peculiar cosmological solutions in modified-gravity theories. We already know that the Lorentz-invariant galileid \eqref{LI_galileid} provides the correct short-distance description of self-accelerating deSitter solutions in DGP \cite{Nicolis:2004qq} and massive gravity \cite{deRham:2010tw}. It would be interesting to embed the other galileids  in modified-gravity theories as well.
In particular, it would be interesting to see whether the multi-galileid with internal $SO(3)$ symmetry discussed in Appendix \ref{general galileids} corresponds to certain cosmological solution in the multi-graviton theory of \cite{Hinterbichler:2012cn}.

We close this section by noticing that, even without invoking modified gravity, there is a general prescription for how to couple the Goldstones of spacetime symmetries to gravity \cite{Delacretaz:2014oxa} in a way that respects all the symmetries. It involves introducing gauge fields for {\it all} the non-linearly realized spacetime symmetries. In the galileon case, this would mean introducing a gauge field for the $Q^\mu$'s as well. Then, like in more standard cases, it is conceivable that one could reduce the number of independent gauge fields by  imposing certain covariant constraints, like the torsion-free one for instance. This procedure would provide another way to couple our galileids to gravity---or better, to an extension of gravity with more  degrees of freedom. In fact, it is entirely possible that this way one would be able to reconstruct certain general features of modified-gravity theories from the bottom-up---perhaps the first non-trivial corrections beyond the decoupling limit.

\subsection{Other extra-ordinary stuff} \label{otherstuff}

Despite the difficulties encountered in realizing  symmetry breaking patterns 3 and 5, we should make clear that these problems do not affect {\it all} extra-ordinary systems---i.e., all those systems whose additional symmetries do not commute with Poincar\'e. As a simple counterexample, we can consider a system with an order parameter made up of three $U(1)$ gauge fields $A_\mu^I$  with a {\it global} $SO(3)$ symmetry that rotates them into each other. (This $U(1)^3$ vs.~$SO(3)$ mismatch is consistent as long as the $A_\mu$'s don't couple to matter). Now, let's imagine that the order parameter acquires an expectation value of the form $\langle A_\mu^I \rangle = \alpha \, \delta_\mu^0 x^I$
\footnote{This corresponds to having three ``electric" fields perpendicular to each other. One could also consider the ``magnetic'' version, in which case the expectation value would be  $\langle A_\mu^I \rangle = \epsilon^{IJK}\delta_{\mu J} x_K$.}. 
This quantity is invariant under a combination of internal shifts and translations, as well as internal and spatial rotations. Such a system---which for lack of a better name we will dub {\it gaugid}---belongs to the same class as ordinary solids (class 7), but entails an infinite number of generators that do not commute with Poincar\'e. Its effective Lagrangian is an arbitrary function of all Lorentz- and $SO(3)$-invariant contractions of the field strength $F_{\mu\nu}^I = \d_\mu A^I_\nu - \d_\nu A^I_\mu$. Consequently, its stress-energy tensor depends on first derivatives of $A^I_\mu$ and is generically  such that $\rho + p \neq 0$ for the background above; furthermore, $\rho$ and $p$ both depend on the thermodynamic control parameter $\alpha$ that appears in the vev of $A_\mu^I$. Finally, the purely free case where $\mathcal{L} = - \fr{1}{4} F_{\mu\nu}^I F^{\mu\nu}_I$ is clearly free of instabilities, and sufficiently small corrections to this Lagrangian will preserve this state of affairs. Thus, as far as we understand, gaugids seem to have all the necessary prerequisites to behave like condensed matter systems. Do they actually exist?

We would also like to mention the intriguing possibility of extra-ordinary systems whose additional spacetime symmetries include supersymmetry (SUSY). More specifically, one could envision a static, homogeneous and isotropic systems that spontaneously break SUSY and some of Poincar\'e generators down to some diagonal linear combinations. We should stress that this SUSY could be completely independent from any SUSY that may or may not play a role in our description of elementary particles. Instead, it should be treated conservatively as a convenient technical device with a limited regime of applicability, exactly like the galileon symmetry studied in sec. \ref{sec:galileids}, or perhaps like the internal symmetries of solids and fluids (more in the next section). The first thing to notice is that the minimal $N=1$ case would not be sufficient to carry out this program.  In fact, in order to use the SUSY generators to define some effective unbroken generators $\bar P_0, \bar P_i$ and $\bar J_i$, we would need to build linear combinations of the supersymmetric charges $Q_\alpha$ and their complex conjugates that transform as ``singlets'' and ``triplets'' under the rotations generated by some suitably defined $\bar J$'s. Basic group-theory considerations are sufficient to realize that this cannot be done for $N=1$ SUSY, because the $Q$'s transform only under the Lorentz group and they are already in an irreducible spin-1/2 representation. This ceases to be an obstacle already for $N=2$, in which case the Lorentz and $R$-symmetry indices of $Q_\alpha^a$ can be contracted to yield a complex singlet, $\delta^\alpha_a Q_\alpha{}^a$, and a complex triplet, $(\sigma^i)_a{}^\alpha Q_\alpha{}^a$, under a combination of spatial rotations and $R$-symmetry transformations. The major obstacle to carrying out this program is that the thermodynamic parameters would need to be Grassmann variables in order for the unbroken generators to have definite Grassmann parity. The physical interpretation of such thermodynamic parameters would be far from clear, and at least for the moment this discourages us from further exploring this possibility.

%%%%%%%%%%%%%%%%%%%%%%%%%%%%%%%%%%%%%%%%%%%%
%%%%%%%%%%%%%%%%%%%%%%%%%%%%%%%%%%%%%%%%%%%%

\section{Summary and discussion}

In this paper we have  taken the spontaneous breaking of Lorentz boosts as the  defining property of condensed matter. 
For a field like condensed matter physics, which has witnessed such exquisitely concrete experimental and theoretical breakthroughs, our abstract approach based purely on symmetry arguments
% and deliberately oblivious to microphysical details 
might appear  both temerarious and naive.
%
%For a field, condensed matter physics, that is witnessing such amazing---and exquisitely concrete---breakthroughs, our abstract approach might appear both temerarious and ingenuous. 
At the same time, the generality of this approach has led us to interesting results, which we summarize in the following points and interleave with comments and open questions.

\begin{itemize}
\item[(1)] Relativistic (Poincar\'e-invariant) field theories, possibly supplemented with additional symmetries, admit {\it eight} inequivalent ways in which Lorentz boosts can be  broken by states that preserve homogeneity and isotropy (sect.~\ref{SSB}). Each of these states can be seen as the ground state of a distinct condensed matter system.  
\end{itemize}
Although our focus here is on condensed matter, the spontaneous breaking of boosts also occurs in cosmology, where the very presence of the cosmic microwave background gives rise to a preferred frame. We expect therefore that some of our symmetry breaking patterns might find interesting applications there too
\footnote{See e.g.~\cite{Jacobson:2000xp} and~\cite{ArkaniHamed:2003uy} for modifications of gravity based on type-I framids and superfluids respectively, or~\cite{Gruzinov:2004ty, Endlich:2012pz, Sitwell:2013kza} for an inflationary model based on solids.}.
\begin{itemize}
\item[(2)]  The role of the additional symmetries is to combine with the Poincar\'e generators to yield new unbroken translations ($\bar P^i$, $\bar P^0$) or  rotations ($\bar J^i$). Six out of our eight scenarios can be realized by supplementing the Poincar\'e group with purely {\it internal} symmetries. The remaining two need instead additional symmetries that do not commute with Poincar\'e (see Table~\ref{t1}).  
\end{itemize}
What is the physical meaning of such additional symmetries?  In the case of superfluids, they can be traced to the microscopic theory.
% at least as approximate low energy features.
For {type-I superfluids}, the internal $U(1)$ symmetry is associated with the conservation of  some ``particle number", be it baryon number for superfluid phases of QCD, or helium-atom number for superfluid helium-4. 
%{\color{blue} Da qui in poi ho tagliato moltissimo: confrontate con le versioni precedenti per favore, vedete se vi sta bene.} 
For  superfluid helium-3 (the most concrete realization of a {type-II superfluid}), there is a similar story, whereby on top of the $U(1)$ symmetry generated by $Q$ there is a spin $SO(3)$ symmetry, which in the approximation of vanishing spin-orbit couplings can be taken as independent from the orbital one, and can thus be thought of as ``internal".

However, the case of ordinary solids and fluids is much less clear. There seems to be no trace of their internal symmetries in the microscopic theory. These are internal 3D shifts and rotations for isotropic solids, and internal 3D volume-preserving diffeomorphisms for fluids. Clearly, these symmetries are just not there in particle physics,  even approximately. Nor would we content ourselves with their appearing as approximate symmetries only: in the low-energy EFT, they are supposed to be valid to all orders in the derivative expansion---they are supposed to be {\it exact} symmetries.
Given that these symmetries seem neither fundamental nor accidental, it appears that the only remaining option is to interpret them as  gauge redundancies.
For that option to make sense we should be able to describe the dynamics of these systems by purely using invariant (but possibly non-local) operators. One can check that this is indeed the case. For instance, in the case of fluids, that simply amounts to working with Eulerian coordinates. In the quantum description of the fluid field theory the gauging of the internal volume preserving diffs also eliminates a virtually infinite degeneracy of each energy level \cite{Endlich_toappear}. However it is not very clear to us how to trace the origin of this redundancy. Should it be associated  with the quantum indistinguishability of the atoms that make up such systems? But if that is the case, why don't we have a more extended permutation symmetry  in the case of the solids?
This question is certainly not settled, and we think it deserves further study. 
\begin{itemize}
\item[(3)]  Only four out of our eight scenarios are experimentally known so far.  In this paper, we set out to study the remaining four. In order to discuss their most robust and model independent features, we have restricted our attention to their Goldstone sector.
\end{itemize}
Why doesn't our classification  exhaust all known existing condensed matter systems? 
By focusing on the Goldstone bosons from the start, we have given up recovering fermionic excitations, even the gapless ones. 
However, systems like Fermi liquids~\cite{Landau_statistics} evade our general classification also as far as their gapless {\it bosonic} excitations---zero sound and spin waves---are concerned. Given that we do not fully understand the physical origin of the solid and fluid internal symmetries, it is conceivable that new symmetries are needed to recover the gapless excitations of Fermi liquids. In fact, an intriguing possibility is that gapless {\it fermionic} excitations emerge as {\it goldstinos}, that is, as Goldstone excitations associated with the spontaneous breaking of a fictitious SUSY, along the lines discussed in sect.~\ref{otherstuff}.
%In particular, we do not seem to be able to reproduce fermionic excitations in a straightforward way. Of course, fermions can always be attached to the picture as additional ``matter fields", but they do not seem to be  needed on the basis of symmetry arguments. For instance, systems like Fermi liquids~\cite{Landau_statistics}, with spin-$1/2$ excitations, evade our general classification. 
%%
%%
\begin{itemize}
\item[(4)]  The simplest possible scenario---the one that does not involve any additional symmetry---yields a perfectly sensible effective theory for the three Goldstone modes associated with the broken boosts (sect.~\ref{sec-3}). This system, which we dubbed {\it type-I framid}, does not have some of the usual properties of condensed matter. For instance, its ground state features an ultra-relativistic stress-energy tensor with $\rho + p =0$. Moreover, there  are no thermodynamic control parameters that can continuously vary its energy density and pressure (sect.~\ref{why not}). 
\end{itemize}
These arguments provide empirical reasons for the non-existence of framids: such systems just happen to behave very differently than all the objects around us. 
The deeper question remains, however, of why the symmetry breaking pattern of framids is not realized in the real world, or, equivalently, why the Goldstones of Lorentz boosts never seem to appear in nature. We will reiterate on this question at the end of this section.  
%%
%5
\begin{itemize}
\item[(5)] A close relative of the { type-I framid}, sharing all its basic properties, is the {type-II framid}. Its low energy theory also contains the three Goldstones for the broken boosts (this is the defining property of ``framids" in our notation), but also three more Goldstones for the broken rotations (sect.~\ref{sec-3}). 
\end{itemize}
Even though some features of the framids are not welcome from a condensed matter perspective, they can be appealing in a cosmological context. For instance, like the ghost condensate \cite{ArkaniHamed:2003uy}, framids have an equilibrium stress-energy tensor that is on the verge of violating the null energy condition (NEC). At the same time, they spontaneously break some Poincar\'e symmetries, which implies that fluctuations in the stress-energy tensor start at {\it linear} order in the Goldstone fields, that is, they have no definite sign. Therefore, roughly speaking, framids violate the NEC $50\%$ of the time
\footnote{Indeed, some of our colleagues may consider this to be {\it the} reason why framids do not seem to be realized in nature.}---and they do so while featuring a perfectly well-behaved (stable, sub-luminal) spectrum of excitations
%even though the corresponding backgrounds are not homogeneous and isotropic.
\footnote{The NEC violation for the particular case of spherically symmetric backgrounds was already discussed in~\cite{Eling:2006df}.}. 
More in general, we believe that our framework provides an ideal starting point to extend the results of \cite{Dubovsky:2005xd} on possible NEC violations by physically well-behaved systems.
\begin{itemize}
\item[(6)] The scenarios corresponding to cases 3 and 5 of Table~\ref{t1} necessarily require additional spacetime symmetries. In sect.~\ref{sec:galileids}  we have studied their simplest realizations: with the smallest number of symmetry generators (four), or  the smallest number of Goldstones (one). We have found that these systems cannot be sensible condensed matter systems, because either they are plagued by instabilities, or the expectation value of their stress-energy tensor is not homogeneous. 
\end{itemize}
It remains unclear whether these pathologies affect all the systems in these two classes, or only the ones we have considered. For sure, these problems do not affect all {\it extra-ordinary systems}---i.e., all those systems whose additional symmetries do not commute with Poincar\'e. As a simple counterexample, we have mentioned the ``{gaugids}" in sect.~\ref{otherstuff}.

Finally, after all the emphasis that we have been giving to the breaking of Lorentz boosts as the defining property of the objects around us, quite ironically, we are left with the puzzle of why the corresponding Goldstone fields (the ``framons") never appear to be there. The inverse Higgs mechanism (sect.~\ref{SSB}) gets rid of the framons in all four scenarios realized in nature.
However, in order for the inverse Higgs mechanism to apply, we need to be in the presence of suitable (broken) symmetries. In all existing systems there seems to be enough additional symmetries for this to happen, although we do not always understand their micro-physical origin. Why is this such a widespread  feature of condensed matter?  Why is nature never showing the Goldstones of the Lorentz boosts?

%%%%%%%%%%%%%%%%%%%%%%%%%%%%%%%%%%%%%%%%%%%%
%%%%%%%%%%%%%%%%%%%%%%%%%%%%%%%%%%%%%%%%%%%%
\section*{Acknowledgements}
\addcontentsline{toc}{section}{Acknowledgements}

We are particularly indebted to Rachel A. Rosen for many enlightening discussions and for collaboration in the early stages of this project. We also would like to thank Nima Arkani-Hamed, Paolo Creminelli, Solomon Endlich, Raphael Flauger, Daniel Green, Ted Jacobson, Dam Thanh Son for interesting discussions and valuable feedback.
FP acknowledges the financial support of the UnivEarthS Labex program at Sorbonne Paris Cit (ANR-10-LABX-0023 and ANR-11-IDEX-0005-02),
of the DOE under contract DE-FG02-11ER41743 and of the A*MIDEX project (n¡ ANR-11-IDEX-0001-02) funded by the ``Investissements d'Avenir" French Government program, managed by the French National Research Agency (ANR). 
AN and RP are supported by the DOE under Contract No. DE-FG02-11ER41743.
RR is supported by the Swiss National Science Foundation under grant 200020-150060.

\appendix

%%%%%%%%%%%%%%%%%%%%%%%%%%%%%%%%%%%%%%%%%%%%
%%%%%%%%%%%%%%%%%%%%%%%%%%%%%%%%%%%%%%%%%%%%

\section{Energy-momentum tensor of type-I framids} \label{app_B}

In order to compute the energy-momentum tensor of type-I framids we resort to their most straightforward realization: a vector field $A_\mu(x)$ with unit norm that acquires a time-like  expectation value. At second order in derivatives, the action $S = \int d^4x \sqrt{-g} {\cal L}$ can only contain the following terms
\begin{equation}  \label{B1}
{S} \, = \, \int d^4 x \sqrt{-g} \left[c_1 \nabla_\mu A_\nu \nabla^\mu A^\nu + c_2 (\nabla_\mu A^\mu)^2 + c_3 \nabla_\mu A_\nu \nabla^\nu A^\mu + c_4 \dot A_\mu \dot A^\mu \right]\, , 
\end{equation}
where  we indicate with a nabla covariant derivation, and with a
dot derivation along the $A_\mu$ direction, say, $\dot{f} \equiv A^\nu \nabla_\nu f$. Incidentally, as already noted, the above action is nothing but that  of an Einstein-aether theory~\cite{Jacobson:2000xp}, once the unit-norm condition for $A_\mu$ is suitably imposed. Notice that in flat space the $c_2$ and $c_3$ terms are equivalent (upon integration by parts), because there we can commute derivatives. Once that is taken into account, the mapping of the $c_i$'s to the $M_i^2$ coefficients of sect.~\ref{sec-3} is obvious,
\be
c_1 \to -\sfrac12 M_2^2 \; ,  \qquad c_2 + c_3 \to -\sfrac12 M_3^2 \; ,  \qquad c_4 \to -\sfrac12 (M_2^2 - M_1^2) \; . 
\ee

The results that we present for the energy momentum tensor are consistent with those already found for Einstein-aether theories ({\it e.g.}~\cite{Jacobson:2004ts,Carroll:2004ai}). However, in our derivation the constraint $A_\mu A^\mu = -1$ is not imposed by the method of Lagrange multipliers, but by using the vierbein formalism. In the presence of a vierbein $e_\mu^{\ a}$, where latin indices $a,b,c\dots$ are Lorentz indices, we can write a unitary vector field as the action of a boost transformation on the vierbein itself,
\begin{equation} \label{amu}
A_\mu(x) = e_\mu^{\ a} (x) \,  A_a(x)= e_\mu^{\ a} (x) \,  \Lambda_a {}^0 (x) \; ,
\end{equation}
where we used that, by definition, a generic  configuration for $A_a(x)$ in the presence of nontrivial Goldstone fields, is just a boosted version of its vacuum expectation value:
\be
A_a(x) =  \Lambda_a {}^b (x) \, \langle A_b \rangle \; , \qquad \langle A_b \rangle =  \delta_b^0 \; .
\ee

In order to derive the equations of motion, when taking the variation of the action with respect to $A_\mu$, what we really want to be freely varying is the boost matrix $\Lambda_a^{\ 0}(x)$. The latter has only three degrees of freedom---{\it e.g.}, the three goldstones $\eta^i$ that we used to parameterize the boost coset in Sec.~\ref{sec-3}.  Straightforward manipulations  give
\begin{align}
\delta S & =  \int d^4 x \, \frac{\delta S}{\delta A_\mu(x)} \, e_\mu^{\ a}(x) \delta \Lambda_a^{\ 0}(x)\, \\
& = \int d^4 x \, \frac{\delta S}{\delta A_\mu(x)} \, e_\mu^{\ a}(x) \Lambda_a^{\ c} (x) \Lambda_{\ c}^b (x)  \delta \Lambda_b^{\ 0}(x)\, \\
& = \int d^4 x \, \frac{\delta S}{\delta A_\mu(x)} \, e_\mu^{\ a}(x) \left[\delta_a^b - \Lambda_a^{\ 0} (x) \Lambda_{\ 0}^b (x) \right]  \delta \Lambda_b^{\ 0}(x)\, \\
& = \int d^4 x \, \frac{\delta S}{\delta A_\mu(x)} \, \left[\delta_\mu^\nu + A_\mu A^\nu \right] e_\nu^{\ a}(x) \delta \Lambda_a^{\ 0}(x)\, .
\end{align}
Since the term in square brackets is a projector,  we obtain that the equations of motions in the presence of the constraint are just the projection of those obtained by freely varying $A_\mu$,
\begin{equation} \label{eoms}
\left(\delta_\mu^\nu + A_\mu A^\nu \right)   \frac{\delta S}{\delta A_\mu} \ = \ 0 \, .
\end{equation}

Analogously, when computing the energy momentum tensor, we  cannot vary the metric $g^{\mu \nu}$ independently of $A_\mu$, or we would fail to satisfy the constraint $A_\mu A_\nu g^{\mu \nu} = -1$. Again, it suffices to write $A_\mu$ as in~\eqref{amu} and simply calculate the overall variation of the action with respect to the vierbein while keeping $A_a(x)$ constant,
\begin{equation}
\delta S = \int d^4 x \sqrt{-g} \ U^\mu_{\ a}(x) \, \delta e_\mu^{\ a}(x) \, .
\end{equation}
The object with mixed indices $U^\mu_{\ a}(x)$, which is related to the energy momentum tensor simply by $T^{\mu \nu} = e^{\nu a} \, U^\mu_{\ a}$ (see {\it e.g.}~\cite{Weinberg_GR}) , receives the usual contribution from the metric $g_{\mu \nu} = e_\mu^{\ a} e_{\nu a}$ {\it and} the contribution from $A_\mu$, because of~\eqref{amu}.
In summary, we get
\begin{equation}
T^{\mu \nu} = \frac{1}{\sqrt{-g}}\left(2 \frac{\delta S}{\delta g_{\mu \nu}} + \frac{\delta S}{\delta A_\mu} A^\nu \right)\, .
\end{equation}
Upon using the equations of motion~\eqref{eoms}, it is easy to show that $T^{\mu \nu}$ is symmetric and equivalent to that obtained in~\cite{Jacobson:2004ts,Carroll:2004ai} with the method of Lagrange multipliers. Note that the energy momentum tensor defined in this way contains the same number of fields $A_\nu$ as the Lagrangian. In particular, we get
\begin{align} 
T^{\mu \nu} & = {\cal L} \, g^{\mu \nu} \label{timunu_curved}\\
& + 2 c_1 \left[ \nabla_\rho A^{(\mu} \nabla^{\nu)} A^\rho -  \nabla_\rho A^\rho \nabla^{(\mu} A^{\nu )} - \nabla^{(\mu} A^{\rho } \, \nabla^{\nu)} A_{\rho} + A^{(\mu} \nabla_\rho \nabla^{\mu )} A^\rho - A^\rho \nabla_\rho \nabla^{(\mu} A^{\nu )} \right] \nonumber \\
&+2 c_2 \left[-g^{\mu \nu} (\nabla_\rho A^\rho)^2 - g^{\mu \nu} A^\rho \nabla_\rho \nabla_\sigma A^\sigma + A^{(\mu} \nabla^{\nu )} \nabla_\rho A^\rho\right] \nonumber \\
&+2 c_3   \left[ \nabla_\rho A^{(\mu} \nabla^\rho A^{\nu )} - \nabla_\rho A^\rho \nabla^{(\mu} A^{\nu )} 
- \nabla^{(\mu} A^{\rho } \, \nabla_{\rho} A^{\nu)} +  A^{(\mu} \Box A^{\mu )} - A^\rho \nabla_\rho \nabla^{(\mu} A^{\nu )} \right] \nonumber \\
&+2 c_4 \left[\dot A^\rho \nabla_\rho A^{(\mu} A^{\nu)} - \dot A^\rho A^{(\mu}  \nabla^{\nu )} A_{\rho} - \dot A^\mu \dot A^\nu - \nabla_\rho A^\rho A^{(\mu} \dot A^{\nu)} + A^\mu A^\nu \nabla_\rho A^\sigma \nabla_\sigma A^\rho \nonumber \right. \\ 
&\ \ \ \ \ \ \left.+ A^\mu A^\nu A^\rho \nabla_\sigma \nabla_\rho A^\sigma - A^\rho A^\sigma \nabla_\rho \nabla _\sigma A^{(\mu } A^{\nu )}\right]  \nonumber 
\end{align}

As a non-trivial check of the above expression, we can make sure that it is conserved on the equations of motion, at least in the flat space limit, where the possibility of commuting partial derivatives simplifies to some extent the cumbersome calculation. Indeed, one can show that
\begin{equation}
\partial_\rho T^{\rho \nu} = \left[\partial_\mu A^\nu - 2 \partial^\nu A_\mu + A^\nu \partial_\mu  -  \delta_\mu^\nu (\partial_\sigma A^\sigma + A^\sigma \partial_\sigma)\right] (\delta^\mu_\sigma + A^\mu A_\sigma) \frac {\delta S}{\delta A_\sigma} \; ,
\end{equation}
which vanishes on the equations of motion~\eqref{eoms}.

%%%%%%%%%%%%%%%%%%%%%%%%%%
%%%%%%%%%%%%%%%%%%%%%%%%%%

\section{Minimal realizations of cases 3 and 5} \la{appa}

In this appendix we try to identify the minimal set of symmetries and Goldstones compatible
with the homogeinity and isotropy requirements of case 3 (see sect.~\ref{SSB}),
\begin{equation}
\bar{P}_0 = P_0, \qquad  \bar{P}_i =P_i+ Q_i, \qquad \bar{J}_i = J_i\, .
\end{equation}
It is immediately clear that 
the commutation relations (\ref{unbroken algebra}) require  
\be \la{3}
[ J_i, Q_j] = i \epsilon_{ijk} Q^k \; ,
\ee
and this implies that the $Q_i$'s cannot generate an internal symmetry. Then, for consistency, these generators must belong to a multiplet that transforms according to  some representation of the Lorentz group. The simplest possibility is the fundamental representation, in which case there must be another symmetry generator $Q_0$ such that the $Q_\mu$'s make up a Lorentz 4-vector. Then we must have
\be \la{s3comm}
[  Q_\mu, Q_\nu ] = 0 \; , \qquad \qquad [  Q_\mu, P_\nu ] = i \eta_{\mu\nu} Y,
\ee
where $Y$ is a central charge---that is, a generator that commutes with all the others---as we now prove.

Because $[ \bar P_i, \bar P_j] = 0$ and because of isotropy, we must have
\be
[Q_i, Q_j] = - 2 i \epsilon_{ijk} X^k, \qquad \qquad [Q_i, P_j ] = i \epsilon_{ijk} X^k + i \delta_{ij} Y + i Z_{ij},
\ee
where $Y, X_k$ and $Z_{ij}$ are some unspecified symmetry generators, and $Z_{ij}$ is symmetric and traceless. The forth generator $Q_0$ completing the multiplet must be such that 
\be
[K_i, Q_j ] = i \delta_{ij} Q_0 \, , \qquad \qquad [K_i, Q_0 ] = i Q_i \, ,
\ee
just because $Q_\mu$ must behave like a 4-vector. If we now apply the Jacobi identity to the generators $P_0, K_i$ and $Q_j$, and use these commutators together with $[P_0, Q_i] = 0$ (which follows from $[\bar P_0, \bar P_i] = 0$) we find that $X^k = Z_{ij} = 0$ and $[Q_0, P_0] = - i Y$. Similarly, if we apply the Jacobi identity to $Q_i, K_j$ and $Q_k$, we find that $[Q_i,Q_0]=0$. This concludes the derivation of the commutation relations (\ref{s3comm}).

Let us now prove that $Y$ must be a central charge. It is easy to realize that $Y$ must be a Lorentz scalar, and thus it must commute with $K_i$ and $J_i$. Then, by using the Jacobi identity for $Q_\mu, Q_\nu, P_\lambda$ and $Q_\mu, P_\nu, P_\lambda$ one shows that $[Q_\mu, Y] =0$ and $[P_\mu, Y] =0$. This concludes the proof that $Y$ is a central charge.

Thus, the smallest symmetry group necessary to implement this scenario is obtained by setting $Y$ to zero and is generated by the four $Q_\mu$'s. If $Q_0$ remains unbroken, the low-energy effective theory will contain the three Goldstone excitations associated with broken $Q_i$.\footnote{The Goldstone of the broken $K_i$ can be eliminated by imposing an inverse Higgs constraint, because $[K_i, \bar P_0] = i (\bar P_i - Q_i)$.} This is however not the most minimal particle spectrum we can have. In fact, in the presence of a {\it broken} central charge $Y$, we can realize this scenario with only one Goldstone boson. This is because, by virtue of the commutators (\ref{s3comm}), one can impose some inverse Higgs constraints and express the Goldstone fields associated with $Q_i$ and (possibly) $Q_0$ in terms of the Goldstone of $Y$. This scenario---that we have dubbed {\it type-I galileid}---is realised for instance when a galileon field takes an expectation value of the form $\phi(x) = A \, |\vec x|^2$, as discussed in sect.~\ref{sec:galileids}.

The analysis of case 5 (see sect~\ref{SSB}),  
\begin{equation}
\bar{P}_0 = P_0 + Q, \qquad  \bar{P}_i =P_i + Q_i, \qquad \bar{J}_i = J_i\, .
\end{equation}
proceeds along the same lines.
Once again, the $Q^k$'s must transform like a 3-vector under rotations, as encoded in eq.~(\ref{3}). The most economical option is to assume that $Q$ and $Q_i$ transform like the components of a 4-vector $Q_\mu$ under Lorentz transformations. Then, because of Lorentz covariance the $Q_\mu$'s must obey the following commutation relations:
\be 
[  Q_\mu, Q_\nu ] = -2 i X_{\mu\nu}, \qquad \qquad [  Q_\mu, P_\nu ] = i X_{\mu\nu} +  i \eta_{\mu\nu} Y,
\ee
where $X_{\mu\nu}$ is an antisymmetric rank-2 tensor of generators. We can strike the best compromise between having fewer Goldstones and adding fewer symmetries by setting $X_{\mu\nu} = 0$ and assuming that $Y$ is broken. This pattern of symmetry breaking requires only one Goldstone and is once again realized by a galileon field, this time with expectation value $ \phi (x) = A \, |\vec x|^2 + B t^2$. This  defines what we called a {\it type-II galileid}.

%%%%%%%%%%%%%%%%%%%%%%%%%%%%%%%
%%%%%%%%%%%%%%%%%%%%%%%%%%%%%%%
\section{Generalizations of galileids}\label{general galileids}

For the classification of  sect.~\ref{SSB}, can we replace some of the internal generators $Q$ studied there with galileon-type  generators $B^\mu$?

The simplest example along these lines is that of a superfluid-like symmetry breaking pattern, 
\be
\bar P^0 = P^ 0 + Q\; , \qquad \bar P^ i = P^i \;, \qquad \bar J = J^i \; , \qquad \mbox{(case 2),}
\ee
which we can achieve through a type-II galileid by choosing $\alpha = 0$, and the corresponding solution for $\beta$:
\be \label{superfluid galileid}
\phi(x) = \sfrac 12  \beta t^2 \; .
\ee
The unbroken time-translations are generated by $\bar P^0 = P^0 + \beta B^0$, and---given the different algebra---the low-energy dynamics for the Goldstone mode will be quite different from those of a standard superfluid's phonon.
In particular, interactions will be substantially softer at low energies, with $2\to2$ scattering amplitudes scaling as $E^6 $ rather than $E^4$.

Similar considerations apply for a solid-like symmetry breaking pattern,
\be
\bar P^0 = P^ 0\; , \qquad \bar P^ i = P^i + Q^i \;, \qquad \bar J^i = J^i  + \tilde Q^i \; , \quad \mbox{(case 7).}
\ee
For this, we need an extension of the galileon algebra with a multiplet of $D$'s and $B$'s transforming non trivially under an internal $SO(3)$ symmetry (generated by the $\tilde Q$'s),
\be
[P^\mu, B^\nu_A] = \eta^{\mu\nu} D_A \; ,\qquad  [B^\mu_A, B^\nu_B] = 0 \;, \qquad \mbox{etc.,}
\ee
where $A$ is an index running over the components of such a representation.
 The simplest possibility would be the vector representation $D_a$, $B_a^\mu$ ($a=1,2,3$). However there is no linear combination of $B_a^\mu$'s that transforms as the vector representation of $\bar J^i = J^i  + \tilde Q^i$, thus making it impossible to mix $P^i$ and $B_a^\mu$ to define unbroken spatial translations with the right algebra with the unbroken rotations. The next possibility is the spin-2 (i.e., symmetric and traceless tensor) representation 
$D_{ab}$, $B_{ab}^\mu$. In that case one could have unbroken combinations of the form
\be
\bar P^0 = P^ 0\; , \qquad \bar P^ i = P^i + T_{ab}^{ij} B_{ab} ^j \;, \qquad \bar J^i = J^i  + \tilde Q^i \; ,
\ee
where the tensor $T$ is defined as
\be
T_{ab}^{ij} = \sfrac12(\delta_a^i \delta_b^j + \delta_a^j \delta_b^i) - \sfrac13 \delta_{ab} \delta^{ij} \; ,
\ee
and makes $Q^i \equiv T_{ab}^{ij} K_{ab} ^j $ transform precisely in the vector representation of $\bar J^i = J^i  + \tilde Q^i$:
\be
[J^i  + \tilde Q^i, Q^j] = i \epsilon^{ijk} \, Q^k \; .
\ee
Then these unbroken combinations have the right algebra for space-time translations and spatial rotations.

These symmetries define a multi-galileon theory \cite{Hinterbichler:2010xn, Padilla:2010de}, with a multiplet of scalars $\phi^{ab}(x)$ transforming as a spin-2 representation of the internal $SO(3)$ symmetry, and as
\be
\phi^{ab}(x) \to \phi^{ab}(x)  + c^{ab} + b^{ab}_\mu \, x^\mu \; , \qquad c^{aa} = b^{aa}_\mu = 0 \; ,
\ee
under the $D$'s and $B$'s.
 The configuration
\be \label{SO(3) galileid}
\phi^{ab}(x) = \sfrac 12 \alpha (x^a x^b - \sfrac13 |\vec x|^2 \, \delta^{ab}) 
\ee
breaks the Poincar\'e group and the internal $SO(3)$ down to translations and rotations generated by the combinations
\be
\bar P^0 = P^ 0\; , \qquad \bar P^ i = P^i + \alpha \, T_{ab}^{ij} B_{ab} ^j \;, \qquad \bar J^i = J^i  + \tilde Q^i \; ,
\ee
as desired. Like for the single galileon case, such a configuration will generically be a solution to the field equations only for discrete choices of the parameter $\alpha$. 
This system features a total of five gapless Goldstone excitations $\pi^{ab}(x) \equiv \delta \phi^{ab}(x)$---the five independent components of a spin-2 representation of $SO(3)$---, which are the minimum number compatible with this symmetry breaking pattern. Out of the 26 broken generators (3 boosts $K^i$, 5 shifts $D_{ab}$, 15 galilean shifts $B^i_{ab}$, and 3 internal rotations $\tilde Q^i$), only the 5 shifts $D_{ab}$ necessarily come with independent Goldstone modes. The others can be non-linearly realized on the same Goldstone fields, thanks to inverse Higgs constraints associated with the commutation relations
\be
[\bar P^0, K^i] = \bar P^i - \alpha T^{ij}_{ab} B^j_{ab} \; , \qquad  [\bar P^i, \tilde Q^k ] = i \epsilon^{jkl} ( B_{li}^j + \delta^j_i B_{lm}^m ) \; , \qquad [\bar P^i, B^j_{ab}] = \delta^{ij} D_{ab} \; .
\ee

We can also combine these two systems into a supersolid-like system (case 8), by considering a reducible representation of the internal $SO(3)$ symmetry---spin-zero and spin-two---both for the $D$ and $K$ generators and for the $\phi$ fields. Then the backgrounds \eqref{superfluid galileid} and \eqref{SO(3) galileid}  preserve unbroken translations and rotations of the form
\be \label{galileid supersolid}
\bar P^0 = P^ 0 + \beta K^0 \; , \qquad \bar P^ i = P^i + \sfrac12 \alpha \, T_{ab}^{ij} K_{ab} ^j \;, \qquad \bar J^i = J^i  + \tilde Q^i \; .
\ee
Unlike for the type-II framid,  the solutions to the field equations with these unbroken symmetries do not form a continuum: $\alpha$ and $\beta$ are uniquely determined up to a finite number of discrete choices, since we have two field equations---one for $\phi$ and one for $\phi^{ab}$---yielding two polynomial equations for $\alpha$ and $\beta$. 

Alternatively, notice that we can also achieve a supersolid-like symmetry breaking pattern through a {\it spin-one} representation $D_a$, $B^\mu_a$ of an internal $SO(3)$:
\be
\bar P^0 = P^0 + \alpha  B^i_i \; , \qquad \bar P^i = P^i - \alpha B^0_i \; , \qquad \bar J^i = J^i  + \tilde Q^i \; .
\ee
These have the right algebra for space-time translations and spatial rotations, and are the symmetries preserved by an $SO(3)$-triplet galileon field on the background
\be
\phi^a(x) = \alpha \, t \, x^a \; ,
\ee
with $\alpha$ being once again a root of the polynomial associated with the field equations.
We have a total of three Goldstone excitations---the independent fluctuations of $\phi^a$. Not only are these considerably fewer than the six associated with the breaking pattern \eqref{galileid supersolid}. They are also fewer than the four associated with a standard supersolid (case 8 in sect.~\ref{SSB}).
 
\newpage  
 
\section{Build your own condensed matter octahedral dice} 

Cut along the perimeter and bend along... you will figure it out:\\
\begin{center}  
    \includegraphics[width=14cm]{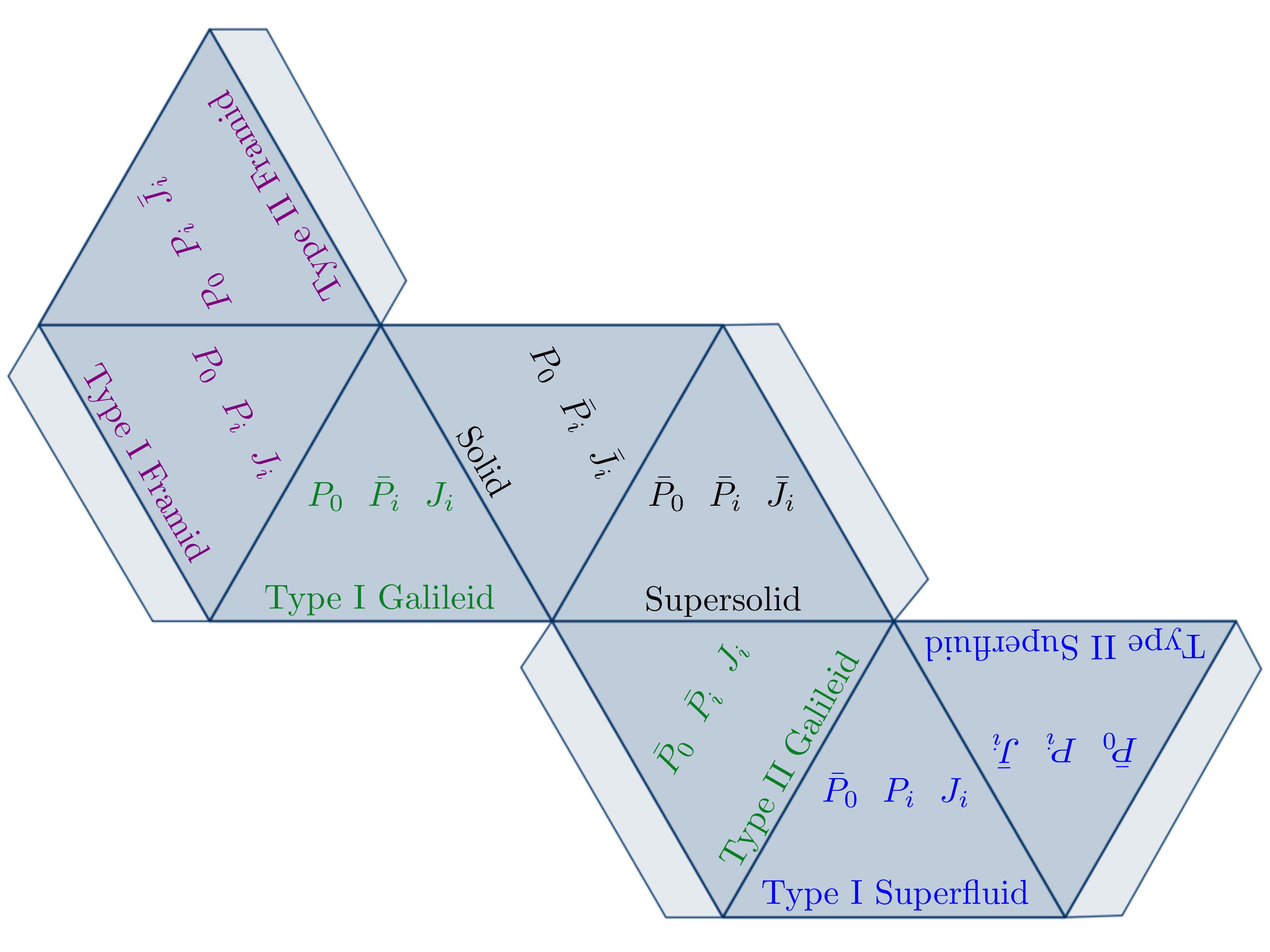}
\end{center}
The dice has some curious properties:
\begin{itemize}
\item {\it Contiguity.} Adjacent faces are algebraically close, in the sense that they differ by the breaking of a single symmetry, be it time translations, spatial translations, or rotations.
\item {\it Complementarity.} Opposite faces are algebraically complementary, in the sense that what is broken for one is unbroken for the other.
\end{itemize}
However, we see no obvious physical counterparts for these properties. For instance, the supersolid and the type-II galileids are contiguous, but their dynamics are clearly very different.

\end{fmffile}

\newpage 

\thispagestyle{empty}
\mbox{}

\newpage
\bibliographystyle{JHEP}
\bibliography{framids}

\end{document}